%% file: ITforRIS.tex
\documentclass[journal,draftclsnofoot,onecolumn,12pt,comsoc, twoside,a4paper]{IEEEtran}

\input{packages_commands_theorems.tex}

\usepackage[noamssymbols]{newtxmath}
\usepackage{eucal}

\usepackage[capitalise]{cleveref}

\usepackage{bm}

\newcommand{\diag}{\mathop{\mathrm{diag}}}

\newcommand{\stack}{\mathop{\mathrm{vec}}}

\newcommand*\diff{\mathop{}\!\mathrm{d}}

\usepackage{acronym}
\acrodef{CSI}[CSI]{channel state information}
\acrodef{CSI2}[CSI]{channel state information}
\acrodef{CSIT}[CSIT]{CSI at the transmitter}
\acrodef{RIS}[RIS]{reconfigurable intelligent surface}
\acrodef{MMSE}[MMSE]{minimum mean-square error}
\acrodef{SNR}[SNR]{signal-to-noise ratio}
\acrodef{ASK}[ASK]{amplitude shift keying}
\acrodef{PSK}[PSK]{phase shift keying}
\acrodef{SCD}[SCD]{successive cancellation decoding}
\acrodef{MIMO}[MIMO]{multiple-input multiple-output}
\acrodef{PSK}[PSK]{phase-shift keying}
\acrodef{CGF}[CGF]{cumulant-generating function}
\acrodef{iid}[i.i.d.]{independent and identically distributed}
\acrodef{QAM}[QAM]{quadrature amplitude modulation}
\acrodef{RF}[RF]{radio frequency}

\begin{document}
	
\bstctlcite{IEEEexample:BSTcontrol} 

\title{Adaptive Coding and Channel Shaping Through
	Reconfigurable Intelligent Surfaces: An Information-Theoretic Analysis}

\author{Roy~Karasik,~\IEEEmembership{Student Member,~IEEE,}
	Osvaldo~Simeone,~\IEEEmembership{Fellow,~IEEE,}
	Marco~Di~Renzo,~\IEEEmembership{Fellow,~IEEE,}
	and~Shlomo~Shamai~(Shitz),~\IEEEmembership{Life Fellow,~IEEE}
\thanks{\looseness=-1 This work has been supported by the European Research Council (ERC) and by the Information and Communication Technologies (ICT) under the European Union’s Horizon 2020 Research and Innovation Programme (Grant Agreement Nos. 694630, 725731, and 871464). This article was presented in part at the 2020 IEEE International Symposium on Information Theory.}
\thanks{R. Karasik and S. Shamai are with the Department of Electrical Engineering, Technion --- Israel Institute of Technology, Haifa 32000, Israel. (royk@campus.technion.ac.il).}
\thanks{O. Simeone is with the Centre for Telecommunications Research,
	Department of Informatics, King’s College London, London WC2R 2LS, U.K.
	(osvaldo.simeone@kcl.ac.uk).}
\thanks{M. Di Renzo is with Universit\'e Paris-Saclay, CNRS, CentraleSup\'elec, Laboratoire des Signaux et Syst\`emes, 3 Rue Joliot-Curie, 91192 Gif-sur-Yvette, France.
	(marco.di-renzo@universite-paris-saclay.fr).}
}

\maketitle

\begin{abstract}
	A communication link aided by a \ac{RIS} is studied in which the transmitter can control the state of the RIS via a finite-rate control link. Channel state information \acused{CSI} (\ac{CSI}) is acquired at the receiver based on pilot-assisted channel estimation, and it may or may not be shared with the transmitter.
	Considering quasi-static fading channels with imperfect \ac{CSI}, capacity-achieving signalling is shown to implement joint encoding of the transmitted signal and of the response of the RIS. This demonstrates the information-theoretic optimality of \ac{RIS}-based modulation, or ``single-RF MIMO'' systems. 
	In addition, a novel signalling strategy based on separate layered encoding that enables practical successive cancellation-type decoding at the receiver is proposed. Numerical experiments show that the conventional scheme that fixes the reflection pattern of the RIS, irrespective of the transmitted information, as to maximize the achievable rate is strictly suboptimal, and is outperformed by the proposed adaptive coding strategies at all practical \ac{SNR} levels.
\end{abstract}

\section{Introduction}
In the context of wireless communications, a \acf{RIS} usually acts as an ``anomalous mirror'' or a ``focusing lens'' that can be configured to reflect or refract impinging radio waves towards arbitrary angles by applying appropriate phase shifts to the incident signals \cite{renzo2020Smart,renzo202Reconfigurable}. Due to these desirable properties, RISs are being considered for future wireless networks as means to shape the wireless propagation channel for signal, interference, security, and scattering engineering 
\cite{wu2020towards,liu2020reconfigurable,renzo2019smart,yuan2020reconfigurable,wu2020intelligent}.

Most prior work, to be reviewed below, proposed to use the RIS as a fixed passive beamformer in order to control the \ac{SNR} levels at the receivers. However, by altering the amplitude or phase of the incident signal, the \ac{RIS} reflection pattern can also be jointly encoded with the transmitted signals as a function of the information message, thus enlarging the modulation space. One instantiation of this idea is the ``single-RF MIMO'' system introduced in \cite{li2020single} that encodes multiple information streams using the \ac{RIS} reflection pattern and a single \ac{RF} chain \cite{tang2020Wireless}. 

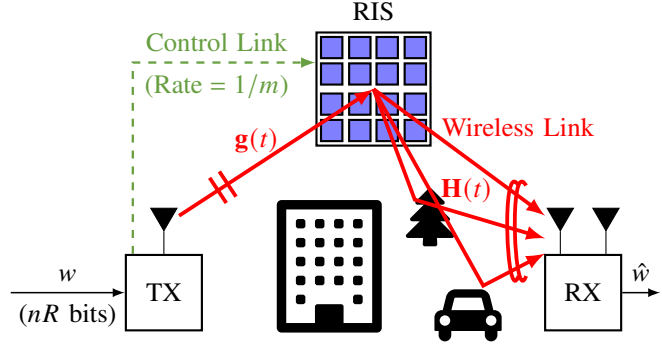
\begin{figure}[!t]
	\centering
	\input{modelFigConf.tex}
	\caption{Illustration of the network under study consisting of a single-\ac{RF} transmitter (TX), a receiver (RX) with $N$ antennas, and an \ac{RIS} with $K$ elements (in the figure, $N=2$ and $K=16$). The transmitter jointly encodes a message $w$ into a codeword of $n$ symbols, sent on the wireless link, and into a control action, sent on the control link to the RIS at a rate of one action every $m$ channel symbols. There is a strong line-of-sight between the transmitter and the RIS, whereas the reflected signal undergoes a multi-path channel.}
	\label{fig:simple-model}
\end{figure}

While practical \ac{RIS}-based modulation schemes exist \cite{tang2020Wireless,basar2019Media,yan2020passive,lin2020reconfigurable,basar2020Reconfigurable,li2020single}, their information-theoretic properties have not been studied. This paper addresses this knowledge gap by studying the capacity of \ac{RIS}-aided communication links in which a single-RF transmitter can control the state of an RIS via a finite-rate control link (see \cref{fig:simple-model}). 
The optimal configuration of the \ac{RIS} requires knowledge of the \ac{CSI}. 
The acquisition of \ac{CSI} is made complicated by the fact that the \ac{RIS} is a nearly-passive device, and hence it cannot process and transmit pilot signals. 
To account for this practical constraint, in this paper, the information-theoretic analysis is based on a model in which the \ac{CSI} is estimated at the receiver via pilot-assisted transmission \cite{jensen2020optimal}, and it may or may not be shared with the transmitter.

\emph{Related Work:}
The optimization of a fixed RIS reflection pattern has been studied in various scenarios. A comprehensive survey of the state-of-the-art is available in \cite{renzo2020Smart}, and we mention here some representative examples. Algorithms for jointly optimizing precoding at the transmitter and beamforming at the RIS were proposed for a point-to-point Multiple-Input Single-Output (MISO) systems in \cite{wu2018Intelligent}, and for Multiple-Input Multiple-Output (MIMO) systems in \cite{perovic2020achievable,zhang2020capacity}. RIS-based passive beamforming was compared to conventional relaying methods such as amplify-and-forward and decode-and-forward in \cite{renzo202Reconfigurable}.

Acquiring \ac{CSI} is crucial for \ac{RIS}-aided communication. 
Channel estimation schemes were proposed in \cite{jensen2020optimal,you2020channel}, in which \ac{RIS} training patterns are designed under the constraint of discrete phase shifts.
The overhead required for channel estimation was studied in \cite{zappone2020overhead}, and an overhead-aware resource allocation framework was developed. Channel estimation based on statistical \ac{CSI} is used in \cite{zhao2020intelligent} to reduce the channel training overhead. 

Schemes for encoding information in the configuration of the \ac{RIS} have been recently presented. In \cite{basar2019Media,yan2020passive,lin2020reconfigurable}, information is encoded in the reflection patterns of the \ac{RIS} by setting the amplitude of each reflecting element to be $0$ or $1$. In \cite{basar2020Reconfigurable}, the receiver antenna for which the \ac{SNR} is maximized encodes the information bits using index modulation \cite{khandani2013media}.
The strategies above are extended in \cite{li2020single} by implementing \ac{PSK} and \ac{QAM} at each element, and by using two independent data streams to control the \ac{RIS}.

\looseness=-1
\emph{Main Contributions:} This work provides an information-theoretic analysis of the RIS-aided system illustrated in \cref{fig:simple-model}, which consists of a single-\ac{RF} transmitter and a receiver with $N$ antennas. \ac{CSI} is assumed to be acquired at the receiver via pilot-based transmission, and it may or may not be shared with the transmitter. We first derive the capacity for any \ac{RIS} control rate, and prove that jointly encoding data onto the transmitted signals and RIS reflection pattern is generally necessary to achieve the maximum information rate.
We explicitly characterize the performance gain of joint encoding in the high-SNR regime.
Then, we propose an achievable scheme based on layered encoding and \ac{SCD} that enables \ac{RIS}-based modulation, while supporting standard separate encoding and decoding strategies. Numerical experiments demonstrate that, for \ac{SNR} levels of practical interest and for a sufficiently fast \ac{RIS} control link, capacity-achieving joint encoding provides significant gain over the max-SNR approach, which fixes the reflection pattern. However, joint encoding is shown to require a more accurate channel estimation compared to the max-SNR scheme, and is hence mostly desirable for long channel coherence blocks. The results in this paper were partially presented in \cite{karasik2020ISIT}, which only considers perfect \ac{CSI} at the transmitter and receiver.

\emph{Organization:} The rest of the paper is organized as follows. In \cref{sec:model}, we present an information-theoretic model for an \ac{RIS}-aided quasi-static fading channel with imperfect \ac{CSI} obtained via channel estimation. In \cref{sec:capacity}, we derive the capacity and we compare it to the rates achieved by two standard suboptimal signalling schemes: a max-SNR scheme that does not encode information in the \ac{RIS} reflection pattern, and an \ac{RIS}-based signalling scheme that modulates the reflection pattern uniformly and has no beamforming gain. In \cref{sec:layered}, we describe an achievable strategy based on layered encoding and successive cancellation decoding with basic separate encoding and decoding procedures. In \cref{sec:bounds}, lower bounds on the capacity and achievable rates are derived. In \cref{sec:numerical}, we present numerical results in order to compare the capacity with the rates achieved by the suboptimal strategies, and to asses the impact of imperfect \ac{CSI} on performance. Finally, in \cref{sec:conclusions}, we conclude the paper and highlight some open problems. 

\emph{Notation:} 
Random variables, vectors, and matrices are denoted by lowercase, boldface lowercase, and boldface uppercase Roman-font letters, respectively. Realizations of random variables, vectors, and matrices are denoted by lowercase, boldface lowercase, and boldface uppercase italic-font letters, respectively. For example, $x$ is a realization of random variable $\mathrm{x}$, $\bm{x}$ is a realization of random vector $\mathbf{x}$, and $\bm{X}$ is a realization of random matrix $\mathbf{X}$.
For any positive integer $K$, we define the set $[K]\triangleq \{1,2,\ldots,K\}$. 
The cardinality of a set $\mathcal A$ is denoted as $|\mathcal{A}|$. 
The Mahalanobis norm of vector $\bm{v}$ with positive semi-definite matrix $\bm{S}$ is defined as $\lVert\bm{v}\rVert_{\bm{S}}\triangleq \sqrt{\bm{v}^*\bm{S}^{-1}\bm{v}}$, where $\bm{v}^*$ denotes the conjugate transpose of vector $\bm{v}$, and the $\ell^2$-norm of a vector $\bm{v}$ is denoted as $\lVert\bm{v}\rVert$. 
$\diag(\bm{x})$ represents a diagonal matrix with diagonal given by the vector $\bm{x}$. 
The trace of a matrix $\bm{X}$ is denote as $\tr(\bm{X})$. The vectorization of matrix $\bm{H}$, i.e., the operator that stacks the columns of $\bm{H}$ on top of one another, is denoted by $\stack(\bm{H})$. The Kronecker product of matrices $\bm{A}$ and $\bm{B}$ is denoted by $\bm{A}\otimes\bm{B}$.

\section{System Model}\label{sec:model}
We consider the system depicted in \cref{fig:simple-model}
in which a single-\ac{RF} transmitter communicates with a receiver equipped with $N$ antennas over a quasi-static fading channel in the presence of an \ac{RIS} that comprises $K$ nearly-passive reconfigurable elements. The $K$ reconfigurable elements are spaced half of the wavelength apart, so that the mutual coupling or channel correlation effects can be ignored as a first-order approximation \cite{gradoni2020end}.
We explore the potential improvement in capacity that can be obtained when the transmitter can encode its message $w\in[2^{nR}]$ of rate $R$ [bits/symbol] not only into a codeword of $n$ symbols sent on the wireless link to the receiver, but also in the reflection pattern of the \ac{RIS}. The reflection pattern is controlled through a rate-limited control link, and is defined by the phase shifts that each of the $K$ \ac{RIS} elements applies to the impinging wireless signal.

\begin{figure}[!t]
	\centering
	\input{timeScales.tex}
	\caption{Illustration of a coding slot. Each slot consists of $n/T$ coherence blocks, which, due to the RIS control link rate, contain $\ell$ sub-blocks of $m$ symbols each. }
	\label{fig:times-scales}
\end{figure}
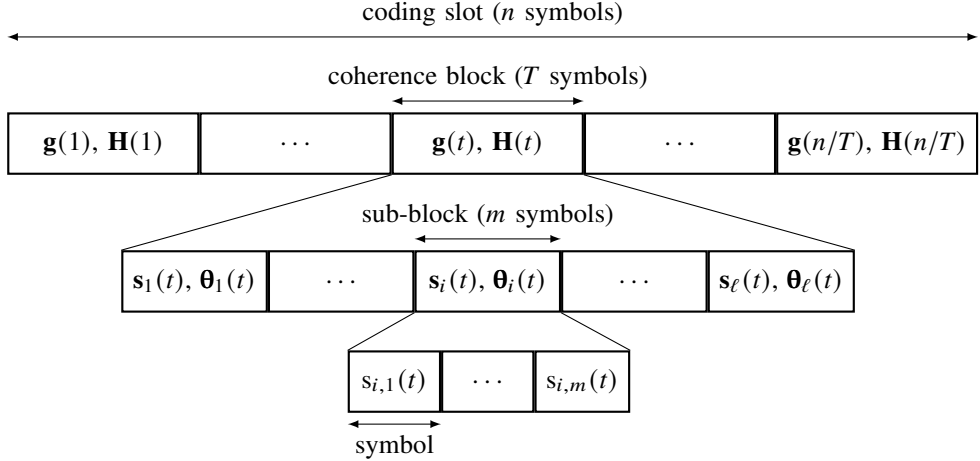

As illustrated in \cref{fig:times-scales}, the fading coefficients are assumed to remain constant for a coherence interval of $T$ symbol periods, after which they change to new independent values. 
The coding slot of $n$ symbols hence contains $n/T$ coherence blocks, which is taken to be an integer.
The codeword transmitted in a coding slot has $n$ symbols from a constellation $\mathcal{S}$ of $S=|\mathcal S|$ points. 
The constellation $\mathcal S$ is assumed to have an average power of one, i.e.,
\begin{IEEEeqnarray}{c}
	\frac{1}{S}\sum_{s\in\mathcal S}|s|^2=1.
\end{IEEEeqnarray} 

The phase shift applied by each element of the \ac{RIS} is chosen from a finite set $\mathcal{A}$ of $A=2^a=|\mathcal{A}|$ distinct hardware-determined values. The \ac{RIS} is controlled by the transmitter by selecting the $K$ phases of the elements as a function of the message $w$. Due to practical limitations on the \ac{RIS} configuration rate, we assume that the phase shifts can only be modified once for each \emph{sub-block} that comprises $m$ consecutive transmitted symbols. As illustrated in \cref{fig:times-scales}, we assume that each coherence block contains $\ell=T/m$ sub-blocks for some integer $\ell\geq 1$, i.e., the \ac{RIS} can be configured at the beginning of each sub-block $i\in[\ell]$ of $m$ transmitted symbols.
Note that if $\ell=1$, i.e., if $m=T$, the reflection pattern of the \ac{RIS} is fixed for the entire coherence block. 

The channel from the transmitter to the \ac{RIS} in the $t$th coherence block, $t\in[n/T]$, is denoted by the vector $\mathbf{g}(t)\in\mathbb C^{K\times 1}$, and the channel from the \ac{RIS} to the $N$ receiving antennas is denoted by the matrix $\mathbf{H}(t)\in\mathbb C^{N\times K}$. 
In order to support multiple information streams with a single \ac{RF} chain, the transmitter and \ac{RIS} are expected to be placed such that there is a strong line-of-sight between them \cite{basar2020Reconfigurable,tang2020Wireless}. Therefore, we assume that the elements of the channel vector $\mathbf{g}(t)$ have random phases and unit amplitude, as illustrated in \cref{fig:simple-model}. In contrast, the reflected signal is assumed to undergo a multi-path channel before being received, and hence the elements of the matrix $\mathbf{H}(t)$ are \ac{iid} as $\mathcal{CN}(0,1)$.
Moreover, as in, e.g., \cite{li2020single,basar2020Reconfigurable}, we assume that the direct link between transmitter and receiver is blocked, so that the propagation from transmitter to receiver occurs solely through the reflected signal from the \ac{RIS}.
During the $t$th coherence block, the fraction of the codeword consisting of $m$ symbols transmitted in the $i$th sub-block, $i\in[\ell]$, is denoted by $\mathbf{s}_i(t)=(\mathrm{s}_{i,1}(t),\ldots,\mathrm{s}_{i,m}(t))^\intercal\in\mathcal S^{m\times 1}$, and is assumed to satisfy
\begin{IEEEeqnarray}{c}
	\frac{1}{m}\mathbb E[\mathbf{s}^*_i(t)\mathbf{s}_i(t)]\leq 1.
\end{IEEEeqnarray}
The phase shifts applied by the \ac{RIS} in the $i$th sub-block are denoted by the vector
\begin{IEEEeqnarray}{c}
	e^{j\pmb{\uptheta}_i(t)}\triangleq (e^{j\uptheta_{i,1}(t)},\ldots,e^{j\uptheta_{i,K}(t)})^\intercal
\end{IEEEeqnarray}
with $\uptheta_{i,k}(t)\in\mathcal{A}$ being the phase shift applied by the $k$th \ac{RIS} element, $k\in[K]$.
Finally, we denote the signal received by the $N$ antennas for the $q$th transmitted symbol by $\mathbf{y}_{i,q}(t)\in\mathbb C^{N\times 1}$, $q\in[m]$. The overall received signal matrix $\mathbf{Y}_i(t)=(\mathbf{y}_{i,1}(t),\ldots,\mathbf{y}_{i,m}(t))\in\mathbb C^{N\times m}$ in the $i$th sub-block can hence be written as
\begin{IEEEeqnarray}{rCl}\label{eq:channel}
	\mathbf{Y}_i(t)&=&\mathbf{H}(t)\diag\left(e^{j\pmb{\uptheta}_i(t)}\right)\mathbf{g}(t)\gamma_i(t)\mathbf{s}^\intercal_i(t)+\mathbf{Z}_i(t)\IEEEnonumber\\
	&=&\bar{\mathbf{H}}(t)e^{j\pmb{\uptheta}_i(t)}\gamma_i(t)\mathbf{s}^\intercal_i(t)+\mathbf{Z}_i(t),
\end{IEEEeqnarray}
where the matrix $\bar{\mathbf{H}}(t)\triangleq \mathbf{H}(t)\diag(\mathbf{g}(t))$, whose elements are \ac{iid} $\mathcal{CN}(0,1)$, combines the channels $\mathbf{g}(t)$ and $\mathbf{H}(t)$; the scalar $\gamma_i(t)>0$ denotes the power gain applied to the transmitted signal $\mathbf{s}_i(t)$, which is subject to the power constraint
\begin{IEEEeqnarray}{c}\label{eq:power_constraint}
	\frac{1}{\ell}\sum_{i=1}^{\ell}\gamma_i^2(t)= P
\end{IEEEeqnarray}
for some $P>0$; 
and the matrix $\mathbf{Z}_i(t)\in\mathbb C^{N\times m}$, whose elements are \ac{iid} as $\mathcal{CN}(0,1)$, denotes the additive white Gaussian noise at the receiving antennas.
It is worth noting that the product $\bar{\mathbf{H}}(t)e^{j\pmb{\uptheta}_i(t)}$ in \eqref{eq:channel} can be viewed as an augmented channel, shaped by the \ac{RIS} for increasing the capacity.

Since the message $w$ is encoded onto both transmitted symbols $\mathbf{s}_i(t)$ and phase shifts $\pmb{\uptheta}_i(t)$, $i\in[\ell]$, $t\in[n/T]$, we denote the effective channel input as
\begin{IEEEeqnarray}{c}\label{eq:def_tilde_X}
	\bar{\mathbf{X}}_i(t)\triangleq \exp\{j\pmb{\uptheta}_i(t)\}\mathbf{s}^\intercal_i(t).
\end{IEEEeqnarray}
With this notation, the channel~\eqref{eq:channel} can be restated as
\begin{IEEEeqnarray}{rCl}\label{eq:equiv_ch}
	\mathbf{Y}_i(t)&=&\gamma_i(t)\bar{\mathbf{H}}(t) \bar{\mathbf{X}}_i(t)+\mathbf{Z}_i(t).
\end{IEEEeqnarray}

\looseness=-1
At first glance, the channel \eqref{eq:equiv_ch} resembles a standard multiple-antenna wireless communication link \cite{hassibi2003How}. In \eqref{eq:equiv_ch}, however, the input matrix $\bar{\mathbf{X}}_i(t)$ is rank-one and is chosen from the finite set 
\begin{IEEEeqnarray}{c}\label{eq:set_C}
	\mathcal C\triangleq \cb{\tilde{\bm{X}}:\tilde{\bm{X}}=\rb{e^{j\theta_1},\ldots,e^{j\theta_K}}^\intercal\bm{s}^\intercal,~ \bm{s}\in\mathcal S^{m\times 1},~\pmb{\theta}\in\mathcal{A}^{K\times 1}}.
\end{IEEEeqnarray}
As a special case, for a fixed \ac{RIS} reflection pattern $\pmb{\uptheta}_i=\pmb{\uptheta}$ for all $i\in[\ell]$, i.e., when the same phase shift vector is used for the entire coherence block, the channel input is chosen from the subset
\begin{IEEEeqnarray}{c}\label{eq:def_subset_C}
	\mathcal C(\pmb{\theta})\triangleq \cb{\tilde{\bm{X}}:\tilde{\bm{X}}=\rb{e^{j\theta_1},\ldots,e^{j\theta_K}}^\intercal\bm{s}^\intercal,~ \bm{s}\in\mathcal S^{m\times 1}}.
\end{IEEEeqnarray}

In the present paper, we study the impact of imperfect \ac{CSI} on the achievable rates. In order to characterize the joint distribution of channel estimation and output signal, we vectorize the channel matrix $\bar{\mathbf{H}}(t)$ and output $\mathbf{Y}_i(t)$ in \eqref{eq:equiv_ch} as
\begin{IEEEeqnarray}{c}\label{eq:vec_ch_coeff}
	\bar{\mathbf{h}}(t)\triangleq\stack(\bar{\mathbf{H}}(t))
\end{IEEEeqnarray}
and
\begin{IEEEeqnarray}{c}\label{eq:vec_equiv_ch}
	\mathbf{y}_i(t)\triangleq\stack(\mathbf{Y}_i(t))=\gamma_i(t)\bar{\mathbf{X}}_i^{\otimes}(t)\bar{\mathbf{h}}(t)+\mathbf{z}_i(t),
\end{IEEEeqnarray}
respectively, where we have defined the vector $\mathbf{z}_i(t)\triangleq\stack(\mathbf{Z}_i(t))\in\mathbb{C}^{Nm\times 1}$, and,
for any matrix $\bar{\mathbf{X}}$, the matrix $\bar{\mathbf{X}}^{\otimes}$ is defined as the Kronecker product
\begin{IEEEeqnarray}{c}\label{eq:mat_A_def}
	\bar{\mathbf{X}}^{\otimes}\triangleq \bar{\mathbf{X}}^\intercal\otimes\bm{I}_N.
\end{IEEEeqnarray}

\subsection{Training and Channel Estimation}\label{sec:training}
\begin{figure}[!t]
	\centering
	\input{training.tex}
	\caption{Structure of a coherence block. The first $\tau$ sub-blocks in each coherence block are used for channel estimation.}
	\label{fig:training}
\end{figure}
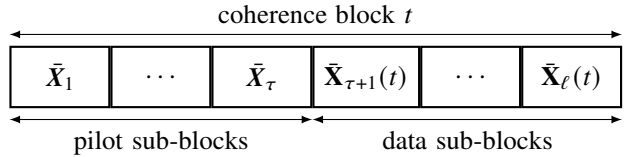

As illustrated in \cref{fig:training}, we focus our attention on transmission schemes in which, for each coherence block $t\in[n/T]$, the first $\tau\geq0$ sub-blocks are used to transmit pilot symbols known to the receiver. That is, we have
\begin{IEEEeqnarray}{c}
	\bar{\mathbf{X}}_i(t)=\bar{\bm{X}}_i,\quad\forall~i\in[\tau],~t\in[n/T],
\end{IEEEeqnarray}
where $\bar{\bm{X}}_1,\ldots,\bar{\bm{X}}_\tau$ denote the pilot symbols. 
The pilot symbols satisfy the power constraint
\begin{IEEEeqnarray}{c}\label{eq:pilots_power_constaint}
	\tr(\bm{X}_{1:\tau}\bm{X}_{1:\tau}^*)\leq Km\tau,
\end{IEEEeqnarray}
where we have defined matrix
\begin{IEEEeqnarray}{c} \label{eq:def_pilots}
	\bm{X}_{1:\tau}\triangleq(\bar{\bm{X}}_1,\ldots,\bar{\bm{X}}_\tau)\in\mathcal C^{1\times\tau}.
\end{IEEEeqnarray}
As for the transmitter, we assume that either it has no access to the \ac{CSI} or that it has access to the receiver's \ac{CSI} via a feedback channel.

The transmission power can vary between the training and information transmission phases. Accordingly, the power gain $\gamma_i(t)$ in \eqref{eq:channel} has two levels
\begin{IEEEeqnarray}{c}
	\gamma_i(t)=\left\lbrace\begin{array}{ll}
		\gamma_\tau&\text{for}~1\leq i\leq\tau,\\
		\gamma_d&\text{for}~\tau+1\leq i\leq\ell.
	\end{array}\right.
\end{IEEEeqnarray}
The power constraint \eqref{eq:power_constraint} can hence be restated as
\begin{IEEEeqnarray}{c}
	\frac{\tau}{\ell}\gamma_\tau^2+\frac{\ell-\tau}{\ell}\gamma_d^2= P.
\end{IEEEeqnarray}
Therefore, the vectorized channel output during the training phase is
 \begin{IEEEeqnarray}{c}\label{eq:train_ch}
 	\mathbf{y}_{1:\tau}(t)\triangleq (\mathbf{y}_1^\intercal(t),\ldots,\mathbf{y}_\tau^\intercal(t))^\intercal=\gamma_\tau\bm{X}_{1:\tau}^{\otimes}\bar{\mathbf{h}}(t)+\mathbf{z}_{1:\tau}(t),
 \end{IEEEeqnarray}
with $\mathbf{z}_{1:\tau}(t)\triangleq(\mathbf{z}_1^\intercal(t),\ldots,\mathbf{z}_\tau^\intercal(t))^\intercal\in\mathbb C^{Nm\tau\times 1}$.

\looseness=-1
Based on the pilot symbols $\bm{X}_{1:\tau}$, the receiver estimates the channel vector $\bar{\mathbf{h}}(t)$ using the \ac{MMSE} estimator, which yields $\hat{\mathbf{h}}(t)=\mathbb E\sqb{\bar{\mathbf{h}}(t)|\mathbf{y}_{1:\tau}(t)}$ as the estimate of $\bar{\mathbf{h}}(t)$ from the observations $\mathbf{y}_{1:\tau}(t)$. Since vectors $\bar{\mathbf{h}}(t)$ and $\mathbf{y}_{1:\tau}(t)$ are jointly Gaussian distributed, the \ac{MMSE} estimator can be computed as the linear MMSE  estimator \cite{kay1993fundamentals}, i.e.,
\begin{IEEEeqnarray}{c}\label{eq:mmse_ch_estimate}
	\hat{\mathbf{h}}(t)=\gamma_\tau(\bm{X}_{1:\tau}^\otimes)^*\rb{\gamma_\tau^2\bm{X}_{1:\tau}^\otimes(\bm{X}_{1:\tau}^\otimes)^*+\bm{I}_{Nm\tau}}^{-1}\mathbf{y}_{1:\tau}(t),
\end{IEEEeqnarray}
and the estimation error is a Gaussian random vector whose covariance matrix is
\begin{IEEEeqnarray}{rCl}\label{eq:gamma_mse_def}
	\mathbf{\Gamma_\text{MMSE}}&\triangleq& \mathbb E\sqb{(\bar{\mathbf{h}}(t)-\hat{\mathbf{h}}(t))(\bar{\mathbf{h}}(t)-\hat{\mathbf{h}}(t))^*}\IEEEnonumber\\
	&=& \bm{I}_{NK}-\gamma_\tau^2(\bm{X}_{1:\tau}^\otimes)^*\rb{\gamma_\tau^2\bm{X}_{1:\tau}^\otimes(\bm{X}_{1:\tau}^\otimes)^*+\bm{I}_{Nm\tau}}^{-1}\bm{X}_{1:\tau}^\otimes.
\end{IEEEeqnarray}

In order to asses how channel estimation affects the achievable performance, we shall also consider as a benchmark the case of perfect \ac{CSI}, which corresponds to the case study in which the vector $\hat{\mathbf{h}}(t)=\bar{\mathbf{h}}(t)$ is available to both the transmitter and receiver as side information without any training ($\tau=0$).
	
\subsection{Channel Encoding}
As discussed, in each coherence block, the transmitter selects the $\ell-\tau$ data sub-blocks
\begin{IEEEeqnarray}{c}\label{eq:def_X}
	\mathbf{X}(t)\triangleq(\bar{\mathbf{X}}_{\tau+1}(t),\ldots,\bar{\mathbf{X}}_\ell(t))\in\mathcal C^{1\times(\ell-\tau)}
\end{IEEEeqnarray} 
based on the information message $w$ and the channel estimate $\hat{\mathbf{h}}(t)$, if available.
The vectorized channel output in \eqref{eq:vec_equiv_ch}, received over the $\ell-\tau$ data sub-blocks, can be expressed as
\begin{IEEEeqnarray}{rCl}\label{eq:vec_data_ch}
	\mathbf{y}(t)\triangleq (\mathbf{y}_{\tau+1}^\intercal(t),\ldots,\mathbf{y}_\ell^\intercal(t))^\intercal &=&\gamma_d\mathbf{X}^\otimes(t)\bar{\mathbf{h}}(t)+\mathbf{z}(t),
\end{IEEEeqnarray}
with $\mathbf{z}(t)\triangleq(\mathbf{z}_{\tau+1}^\intercal(t),\ldots,\mathbf{z}_{\ell}^\intercal(t))^\intercal\in\mathbb C^{Nm(\ell-\tau)\times 1}$.
Having received the vector $\mathbf{y}(t)$ in \eqref{eq:vec_data_ch} for $t\in[n/T]$, the decoder produces the estimate $\hat{w}=\hat{w}(\mathbf{y}(1),\ldots,\mathbf{y}(n/T),\mathcal{H})$ based on the channel estimates $\mathcal{H}\triangleq\{\hat{\mathbf{h}}(1),\ldots,\hat{\mathbf{h}}(n/T)\}$ in \eqref{eq:mmse_ch_estimate}.

For a specific choice of training parameters $\tau$, $\gamma_\tau$, and $\bm{X}_{1:\tau}$, a rate $R(\tau,\gamma_\tau,\bm{X}_{1:\tau})$ is said to be \emph{achievable} if the probability of error satisfies the limit $\Pr(\hat{w}\neq w)\rightarrow 0$ when the codeword length grows large, i.e., $n\rightarrow\infty$. The corresponding ergodic capacity $C(\tau,\gamma_\tau,\bm{X}_{1:\tau})$ is defined as the maximum over all achievable rates, i.e.,
\begin{IEEEeqnarray}{c}\label{eq:cap_def}
	C(\tau,\gamma_\tau,\bm{X}_{1:\tau})\triangleq\sup\{R(\tau,\gamma_\tau,\bm{X}_{1:\tau}):R(\tau,\gamma_\tau,\bm{X}_{1:\tau})\text{ is achievable}\},
\end{IEEEeqnarray}
where the supremum is taken over all joint encoding and decoding schemes. The number of sub-blocks used for training $0\leq\tau\leq\ell$, pilot symbols $\bm{X}_{1:\tau}$, and power-amplifier gain $\gamma_\tau>0$ can all be optimized to increase the achievable rate.

\section{Channel Capacity}\label{sec:capacity}
In this section, we derive the capacity $C(\tau,\gamma_\tau,\bm{X}_{1:\tau})$ defined in \eqref{eq:cap_def} and we prove that the conventional scheme that does not encode information in the RIS reflection pattern 
is strictly suboptimal. More specifically, this result is proved in the high-SNR regime by characterizing the gain of the proposed joint encoding. For finite values of the \ac{SNR}, on the other hand, the performance gain is evaluated in \cref{sec:numerical} via numerical experiments.

Most works on \ac{RIS}-aided systems consider Gaussian codebooks for the transmitted signal $\mathbf{s}_i(t)$. This implies that the resulting achievable rates are formulated in the standard form $\log_2(1+\text{SNR})$, even in the presence of imperfect \ac{CSI} by using standard bounds \cite{zhao2020exploiting}. In contrast, as described in \cref{sec:model}, we focus our attention on the more practical model in which the transmitted symbols and the \ac{RIS} elements' phase response take values from finite sets. 
As a result, standard capacity expressions of the form $\log_2(1+\text{SNR})$ are not applicable, and standard techniques for bounding the capacity under imperfect \ac{CSI} cannot be used.
Specifically, lower bounding the capacity by modeling the residual channel estimation noise as Gaussian \cite{medard2000effect,bustin2014worst} does not hold for finite input constellations \cite{shamai1992worst}. 
Therefore, the expressions for the capacity and achievable rates that we present in this section are more complex, and require the following definitions. 
\begin{definition}
	The \ac{CGF} of a random variable $\mathrm{u}$ is defined as
	\begin{IEEEeqnarray}{c}
		\kappa_r(\mathrm{u})\triangleq \log_2\rb{\mathbb E\sqb{e^{r\mathrm{u}}}},\quad r\in\mathbb R.
	\end{IEEEeqnarray}
	The value of the \ac{CGF} for $r=1$ is denoted by $\kappa(\mathrm{u})\triangleq\kappa_1(\mathrm{u})$.
\end{definition}

\begin{definition}
	The \ac{CGF} of a random variable $\mathrm{u}$ conditioned on a random vector $\mathbf{x}$ is defined as
	\begin{IEEEeqnarray}{c}\label{eq:def_cond_cgf_rand}
		\kappa_r(\mathrm{u}|\mathbf{x})\triangleq \mathbb E\sqb{\log_2\rb{\mathbb E\sqb{e^{r\mathrm{u}}|\mathbf{x}}}},\quad r\in\mathbb R.
	\end{IEEEeqnarray}
	The value of the conditional \ac{CGF} for $r=1$ is denoted by $\kappa(\mathrm{u}|\mathbf{x})\triangleq\kappa_1(\mathrm{u}|\mathbf{x})$.
\end{definition}

We now derive the capacity for the general case with \emph{imperfect \ac{CSI}} available at both the transmitter and receiver. In particular, the capacity is formulated in the form of an optimization problem with respect to the encoding distribution $p_{\mathbf{X}|\hat{\mathbf{h}}}(\bm{X}|\hat{\bm{h}})$ of the effective inputs in \eqref{eq:def_X} given the channel estimate $\hat{\mathbf{h}}$. To this end, we define the covariance matrix of the received signal $\mathbf{y}(t)$ in \eqref{eq:vec_data_ch} conditioned on the channel estimate $\hat{\mathbf{h}}(t)$ and the input $\mathbf{X}(t)$ as
\begin{IEEEeqnarray}{c}
	\mathbb E\sqb{\mathbf{y}(t)\mathbf{y}(t)^*|\hat{\mathbf{h}}(t),\mathbf{X}(t)}=\bm{I}_{Nm(\ell-\tau)}+\gamma_d^2\mathbf{X}^\otimes(t)\mathbf{\Gamma_\text{MMSE}}(\mathbf{X}^\otimes(t))^*=\pmb{\Gamma}(\mathbf{X}(t)),
\end{IEEEeqnarray}
where, for any matrix $\mathbf{X}$, we have defined the positive semidefinite matrix $\pmb{\Gamma}(\mathbf{X})$ as
\begin{IEEEeqnarray}{c}\label{eq:gamma_X_def}
	\pmb{\Gamma}(\mathbf{X})\triangleq \bm{I}_{Nm(\ell-\tau)}+\gamma_d^2\mathbf{X}^\otimes\mathbf{\Gamma_\text{MMSE}}(\mathbf{X}^\otimes)^*.
\end{IEEEeqnarray}
We also define the decomposition
\begin{IEEEeqnarray}{c}\label{eq:mat_V_def}
	\pmb{\Gamma}(\mathbf{X})=\bm{V}(\mathbf{X})\bm{V}(\mathbf{X})^*,
\end{IEEEeqnarray}
where $\bm{V}(\mathbf{X})$ is a square root matrix of $\pmb{\Gamma}(\mathbf{X})$.
\begin{proposition}\label{prop:capacity}
	When the \ac{MMSE} estimate $\hat{\mathbf{h}}(t)$ in \eqref{eq:mmse_ch_estimate} is available at both the receiver and transmitter, the capacity of the channel \eqref{eq:vec_data_ch} is given as
	\begin{IEEEeqnarray}{l}\label{eq:capacity}
		C(\tau,\gamma_\tau,\bm{X}_{1:\tau})
		= -\frac{N(\ell-\tau)}{\ell}\log_2(e)
		-\min_{\substack{
					p_{\mathbf{X}|\hat{\mathbf{h}}}(\bm{X}|\hat{\bm{h}}):\\
					\mathbb E[\tr(\mathbf{X}\mathbf{X}^*)]\leq Km(\ell-\tau),\\
					\mathbf{X}\in\mathcal C^{1\times (\ell-\tau)}}}
			\frac{1}{m\ell}\kappa(\mathrm{u}|\mathbf{X}_1,\mathbf{z},\hat{\mathbf{h}}),
	\end{IEEEeqnarray}
	where random variable $\mathrm{u}$ is defined as
	\begin{IEEEeqnarray}{c}\label{eq:def_u}
		\mathrm{u}\triangleq \ln\rb{\frac{|\pmb{\Gamma}(\mathbf{X}_1)|}{|\pmb{\Gamma}(\mathbf{X}_2)|}}-\left\lVert\bm{V}(\mathbf{X}_1)\mathbf{z}+\gamma_d\rb{\mathbf{X}_1^\otimes-\mathbf{X}_2^\otimes}\hat{\mathbf{h}}\right\rVert^2_{\pmb{\Gamma}(\mathbf{X}_2)}
	\end{IEEEeqnarray}
	with independent random vectors $\mathbf{z}\sim\mathcal{CN}(\mathbf{0},\bm{I}_{Nm(\ell-\tau)})$ and $\hat{\mathbf{h}}\sim\mathcal{CN}(\mathbf{0},\bm{I}_{NK}-\mathbf{\Gamma_\text{MMSE}})$, and random matrices $\mathbf{X}_1,\mathbf{X}_2\sim p_{\mathbf{X}|\hat{\mathbf{h}}}(\bm{X}|\hat{\bm{h}})$ that are conditionally independent given $\hat{\mathbf{h}}$. Furthermore, for $\tau\geq K$, we have the high-SNR limit
	\begin{IEEEeqnarray}{c}\label{eq:capacity_high_snr}
		\lim_{P\rightarrow\infty}C(\tau,\gamma_\tau,\bm{X}_{1:\tau})=\frac{(\ell-\tau)\log_2\rb{|\mathcal C|}}{m\ell},
	\end{IEEEeqnarray} 
	which, for a given cardinality $S=|\mathcal S|$ of the signal constellation, is maximized if the \ac{ASK} modulation is used, i.e.,
	\begin{IEEEeqnarray}{c}\label{eq:ASK_mod}
		\mathcal S=\{\sigma,3\sigma,\ldots,(2S-1)\sigma\},
	\end{IEEEeqnarray}
	where the factor $\sigma\triangleq\sqrt{3/[3+4(S^2-1)]}$ ensures a unit average power constraint. In this case, the high-SNR limit is
	\begin{IEEEeqnarray}{c}\label{eq:lim_high_snr_ASK}
		\lim_{P\rightarrow\infty}C(\tau,\gamma_\tau,\bm{X}_{1:\tau})=\frac{\ell-\tau}{m\ell}\sqb{m\log_2(S)+K\log_2(A)}.
	\end{IEEEeqnarray}	
\end{proposition}
\begin{IEEEproof}
	See Appendix \ref{app_proof_cap}.
\end{IEEEproof}

Achieving the capacity in \eqref{eq:capacity} generally requires joint encoding over the codeword symbols $\mathbf{s}_i(t)$ and RIS reflection variables $\pmb{\uptheta}_i(t)$, for all data sub-blocks $i=\tau+1,\ldots,\ell$, $t\in[n/T]$, as well as joint decoding of the message $w$ at the receiver based on the information encoded over both $\mathbf{s}_i(t)$ and $\pmb{\uptheta}_i(t)$. In \eqref{eq:capacity}, this is specified in the optimization over the distribution $p_{\mathbf{X}|\hat{\mathbf{h}}}(\bm{X}|\hat{\bm{h}})$ of the input $\mathbf{X}(t)=(\bar{\mathbf{X}}_{\tau+1}(t),\ldots,\bar{\mathbf{X}}_{\ell}(t))$ in \eqref{eq:def_X}, which, by \eqref{eq:def_tilde_X}, is a function of both $\mathbf{s}_i(t)$ and $\pmb{\uptheta}_i(t)$. 
However, the high-SNR asymptotic limit in \eqref{eq:capacity_high_snr} implies that, in the high-SNR regime, capacity is achieved by using independent random codebooks with uniform distribution for the codeword symbols $\mathbf{s}$ and the RIS reflection pattern $\pmb{\uptheta}$, and perfect channel estimation can be obtained by using $\tau\geq K$ pilot sub-blocks.

At a computational level, problem \eqref{eq:capacity} is convex (see Appendix \ref{app_proof_cap}), and hence it can be solved by using convex optimization tools. 
Moreover, calculating $\kappa(\mathrm{u}|\mathbf{X}_1,\mathbf{z},\hat{\mathbf{h}})$ in \eqref{eq:capacity} involves evaluating the expectation over the random vectors $\mathbf{z}$ and $\hat{\mathbf{h}}$, and over the random matrices $\mathbf{X}_1$ and $\mathbf{X}_2$. Since $\mathbf{z}$ and $\hat{\mathbf{h}}$ are continuous random vectors, the former expectation may be estimated via an empirical average, while the second requires summing  over $|\mathcal C|^{\ell-\tau}$ terms.

The following two corollaries formulate the capacity under the assumption of imperfect \ac{CSI} available only at the receiver, and under the assumption of perfect \ac{CSI} available at both the transmitter and receiver, respectively.

\begin{corollary}\label{cor:cap_csir}
	When the \ac{MMSE} estimate $\hat{\mathbf{h}}(t)$ in \eqref{eq:mmse_ch_estimate} is available only at the receiver, the capacity of the channel \eqref{eq:vec_data_ch} is given as
	\begin{IEEEeqnarray}{l}\label{eq:capacity_csir_cgf}
		C_\text{CSIR}(\tau,\gamma_\tau,\bm{X}_{1:\tau})
		= -\frac{N(\ell-\tau)}{\ell}\log_2(e)
		-\frac{1}{m\ell}\kappa(\mathrm{u}|\mathbf{X}_1,\mathbf{z},\hat{\mathbf{h}}),
	\end{IEEEeqnarray}
	where the random variable $\mathrm{u}$ is defined as in \eqref{eq:def_u}
	with independent random vectors $\mathbf{z}\sim\mathcal{CN}(\mathbf{0},\bm{I}_{Nm(\ell-\tau)})$ and $\hat{\mathbf{h}}\sim\mathcal{CN}(\mathbf{0},\bm{I}_{NK}-\mathbf{\Gamma_\text{MMSE}})$, and independent random matrices $\mathbf{X}_1,\mathbf{X}_2\sim p_{\mathbf{X}}(\bm{X})=1/|\mathcal{C}|^{\ell-\tau}$ for all $\bm{X}\in\mathcal C^{1\times (\ell-\tau)}$.
	Furthermore, for $\tau\geq K$, we have the high-SNR limit
	\begin{IEEEeqnarray}{c}
		\lim_{P\rightarrow\infty}C_\text{CSIR}(\tau,\gamma_\tau,\bm{X}_{1:\tau})=\frac{(\ell-\tau)\log_2\rb{|\mathcal C|}}{m\ell}.
	\end{IEEEeqnarray} 
\end{corollary}
\begin{IEEEproof}
	It follows from the proof of \cref{prop:capacity} with the caveat that, since the channel estimate $\hat{\mathbf{h}}$ is available only at the receiver, the optimal input distribution $p_{\mathbf{X}}(\bm{X})$ is uniform. This is because the channel coefficients in vector $\bar{\mathbf{h}}$ \eqref{eq:vec_ch_coeff} have uniformly distributed phases (see \cite[Sec. VII]{kramer2005cooperative}).
\end{IEEEproof}

\looseness=-1
Prior works \cite{tang2020Wireless,basar2019Media,yan2020passive,lin2020reconfigurable,basar2020Reconfigurable,li2020single} have considered RIS-based modulation schemes that modulate the RIS reflection pattern independently from the transmitted symbols. By \cref{cor:cap_csir}, an \ac{RIS}-based modulation scheme with independent and uniformly generated random codebooks for the transmitted symbols and reflection pattern is optimal when the transmitter has no access to \ac{CSI}, and hence it cannot use the \ac{RIS} for beamforming.
Furthermore, since the high-SNR limits in \cref{prop:capacity} and \cref{cor:cap_csir} are equal, the availability of the \ac{CSI} at the transmitter does not increase the capacity in the high-SNR regime. 

\begin{corollary}[\!\!{\cite[Proposition 1]{karasik2020ISIT}}]\label{cor:cap_perfect}
	When perfect \ac{CSI} is available at both the receiver and transmitter, the capacity of the channel \eqref{eq:vec_data_ch} is given as
	\begin{IEEEeqnarray}{rCl}\label{eq:capacity_perfect_csi}
		C_\text{perfect}&=&  -N\log_2(e)
		-\min_{\substack{
					p_{\mathbf{X}|\bar{\mathbf{h}}}(\bm{X}|\bar{\bm{h}}):\\
					\mathbb E[\tr(\mathbf{X}\mathbf{X}^*)]\leq Km\ell,\\
					\mathbf{X}\in\mathcal C^{1\times\ell}}}
			\frac{1}{m\ell}\kappa(\tilde{\mathrm{u}}|\mathbf{X}_1,\mathbf{z},\bar{\mathbf{h}}),
	\end{IEEEeqnarray}
	where the random variable $\tilde{\mathrm{u}}$ is defined as
	\begin{IEEEeqnarray}{c}\label{eq:def_tilde_u}
		\tilde{\mathrm{u}}\triangleq -\left\lVert \mathbf{z}+\gamma_d\rb{\mathbf{X}_1^\otimes-\mathbf{X}_2^\otimes}\bar{\mathbf{h}}\right\rVert^2
	\end{IEEEeqnarray}
	with independent random vectors $\mathbf{z}\sim\mathcal{CN}(\mathbf{0},\bm{I}_{Nm\ell})$, $\bar{\mathbf{h}}\sim\mathcal{CN}(\mathbf{0},\bm{I}_{NK})$, and random matrices $\mathbf{X}_1,\mathbf{X}_2\sim p_{\mathbf{X}|\bar{\mathbf{h}}}(\bm{X}|\bar{\bm{h}})$ that are conditionally independent given $\bar{\mathbf{h}}$. Furthermore, we have the high-SNR limit $\lim_{P\rightarrow\infty}C_\text{perfect}=\log_2(|\mathcal C|)/m$.
\end{corollary}
\begin{IEEEproof}
	\looseness=-1
	It follows from the proof of \cref{prop:capacity} by setting $\tau=0$ and $\pmb{\Gamma}_\text{MMSE}=\mathbf{0}$, since the channel vector $\bar{\mathbf{h}}$ is known to both the receiver and transmitter without requiring any training.
\end{IEEEproof}

\subsection{Max-SNR Approach}
Having observed that achieving the capacity generally requires joint encoding of data over the codeword symbols and the \ac{RIS} reflection pattern, we now consider the standard approach
in which the reflection pattern of the RIS is fixed for all data sub-blocks $i=\tau+1,\ldots,\ell$, of the fading block $t$, irrespective of the message $w$, i.e.,  $\pmb{\uptheta}_i(t)=\pmb{\uptheta}(t)$.
We denote the fixed \ac{RIS} reflection pattern by $\pmb{\theta}(\hat{\bm{h}})$ to emphasize that it is chosen based on the channel estimate $\hat{\bm{h}}$ to maximize the achievable rate, and we have the following result.
\begin{proposition}\label{prop:max_snr}
	When the \ac{MMSE} estimate $\hat{\mathbf{h}}$ in \eqref{eq:mmse_ch_estimate} is available at both the receiver and transmitter, an encoding scheme that selects the phase shift vector $\pmb{\theta}(\hat{\bm{h}})$ as a function of $\hat{\bm{h}}$ achieves the rate
	\begin{IEEEeqnarray}{l}\label{eq:rate_max_snr}
		R_\text{max-SNR}(\tau,\gamma_\tau,\bm{X}_{1:\tau})
		= -\frac{N(\ell-\tau)}{\ell}\log_2(e)
		-\min_{\substack{\pmb{\theta}(\hat{\bm{h}}):\\\pmb{\theta}(\hat{\bm{h}})\in\mathcal{A}^{K\times 1}}}
		\min_{\substack{
									p_{\mathbf{X}|\hat{\mathbf{h}}}(\bm{X}|\hat{\bm{h}}):\\
									\mathbb E[\tr(\mathbf{X}\mathbf{X}^*)]\leq Km(\ell-\tau),\\
									\mathbf{X}\in\mathcal C(\pmb{\theta}(\hat{\bm{h}}))^{1\times (\ell-\tau)}}}
			\frac{1}{m\ell}\kappa(\mathrm{u}|\mathbf{X}_1,\mathbf{z},\hat{\mathbf{h}}),\IEEEeqnarraynumspace
	\end{IEEEeqnarray}
	where the random variable $\mathrm{u}$ is defined as in \eqref{eq:def_u} with independent random vectors $\mathbf{z}\sim\mathcal{CN}(\mathbf{0},\bm{I}_{Nm(\ell-\tau)})$, $\hat{\mathbf{h}}\sim\mathcal{CN}(\mathbf{0},\bm{I}_{NK}-\mathbf{\Gamma_\text{MMSE}})$, and random matrices $\mathbf{X}_1,\mathbf{X}_2\sim p_{\mathbf{X}|\hat{\mathbf{h}}}(\bm{X}|\hat{\bm{h}})$ that are conditionally independent given $\hat{\mathbf{h}}$.
	Furthermore, for $\tau\geq 1$, we have the high-SNR limit
	\begin{IEEEeqnarray}{c}\label{eq:max-snr_high_snr}
		\lim_{P\rightarrow\infty}R_\text{max-SNR}(\tau,\gamma_\tau,\bm{X}_{1:\tau})=\frac{(\ell-\tau)\log_2(S)}{\ell}.
	\end{IEEEeqnarray}
\end{proposition}
\begin{IEEEproof}
	\looseness=-1
	For a fixed \ac{RIS} reflection pattern $\pmb{\uptheta}_i(t)=\pmb{\theta}(\hat{\bm{h}}(t))$ with $i=\tau+1,\ldots,\ell$, the channel input $\mathbf{X}(t)$ in \eqref{eq:vec_data_ch} is restricted to the finite set $\mathcal{C}(\pmb{\theta}(\hat{\bm{h}}(t)))^{1\times (\ell-\tau)}$ in \eqref{eq:def_subset_C}. Therefore, the result follows from \cref{prop:capacity} by restricting the input such that only the codeword symbols vary over the data sub-blocks. In \eqref{eq:rate_max_snr}, this is reflected in the optimization over the distribution $p_{\mathbf{X}|\hat{\mathbf{h}}}(\bm{X}|\hat{\bm{h}})$ with $\mathbf{X}\in\mathcal C(\pmb{\theta}(\hat{\bm{h}}))^{1\times (\ell-\tau)}$, where the \ac{RIS} reflection pattern $\pmb{\theta}(\hat{\bm{h}})$ is fixed.
	In addition, the limit \eqref{eq:max-snr_high_snr} follows from \eqref{eq:capacity_high_snr} since, for any fixed \ac{RIS} reflection pattern $\pmb{\theta}(\hat{\bm{h}})$, we have $|\mathcal C(\pmb{\theta}(\hat{\bm{h}}))|=S^m$.
\end{IEEEproof}

The limit in \eqref{eq:max-snr_high_snr} implies that, in the high-SNR regime, the rate of the max-SNR scheme is limited to $(\ell-\tau)\log_2(S)/\ell$. This is because, in each coherence block, the information data is modulated solely onto the $m(\ell-\tau)$ codeword symbols, which are selected from a constellation $\mathcal{S}$ of $S$ points. 
By comparing \eqref{eq:max-snr_high_snr} with \eqref{eq:capacity_high_snr}, we evince that, for any phase response set $\mathcal A$ of $A$ distinct phases, modulating the \ac{RIS} reflection pattern can be used to increase the achievable rate by additional $ K(\ell-\tau)\log_2(A)/(m\ell)$ bits per symbol as compared to the max-SNR scheme. 
However, note that the max-SNR scheme can achieve the high-SNR rate \eqref{eq:max-snr_high_snr} by fixing the \ac{RIS} reflection pattern irrespective of \ac{CSI} and estimating only the effective channel from the transmitter to the receiver. Therefore, the max-SNR approach requires only $\tau\geq 1$ pilot symbols to achieve the high-SNR limit in \eqref{eq:max-snr_high_snr}, whereas joint encoding achieves the limit in \eqref{eq:capacity_high_snr} with $\tau\geq K$ pilot symbols.
For finite values of the \ac{SNR}, the achievable rate in \eqref{eq:rate_max_snr} can be computed by combining convex optimization tools for the inner minimization problem and global optimization tools for the minimization over the set of discrete phase shifts. The corresponding performance loss is evaluated in \cref{sec:numerical} via numerical experiments. 

\looseness=-1
The rates achieved for imperfect \ac{CSI} available only at the receiver and for perfect \ac{CSI} available at both the transmitter and receiver are given in the following two corollaries, respectively.

\begin{corollary}\label{cor:snr_csir}
	When the \ac{MMSE} estimate $\hat{\mathbf{h}}$ in \eqref{eq:mmse_ch_estimate} is available only at the receiver, a transmission scheme in which the phase shift vector $\pmb{\theta}$ is kept fixed achieves the rate
		\begin{IEEEeqnarray}{l}\label{eq:rate_max_snr_csir}
		R_\text{max-SNR}^\text{CSIR}(\tau,\gamma_\tau,\bm{X}_{1:\tau})
		= -\frac{N(\ell-\tau)}{\ell}\log_2(e)
		-\min_{\pmb{\theta}:\pmb{\theta}\in\mathcal{A}^{K\times 1}}
		\frac{1}{m\ell}\kappa(\mathrm{u}|\mathbf{X}_1,\mathbf{z},\hat{\mathbf{h}}),
	\end{IEEEeqnarray}
 	where the random variable $\mathrm{u}$ is defined as in \eqref{eq:def_u} with independent random vectors $\mathbf{z}\sim\mathcal{CN}(\mathbf{0},\bm{I}_{Nm(\ell-\tau)})$ and $\hat{\mathbf{h}}\sim\mathcal{CN}(\mathbf{0},\bm{I}_{NK}-\mathbf{\Gamma_\text{MMSE}})$, and independent random matrices $\mathbf{X}_1,\mathbf{X}_2\sim p_{\mathbf{X}}(\bm{X})=1/|\mathcal C(\pmb{\theta})|^{\ell-\tau}$ for all $\bm{X}\in \mathcal C(\pmb{\theta})^{1\times (\ell-\tau)}$. Furthermore, for $\tau\geq 1$, we have the high-SNR limit
	\begin{IEEEeqnarray}{c}
		\lim_{P\rightarrow\infty}R_\text{max-SNR}^\text{CSIR}(\tau,\gamma_\tau,\bm{X}_{1:\tau})=\frac{(\ell-\tau)\log_2(S)}{\ell}.
	\end{IEEEeqnarray} 
\end{corollary}
\begin{IEEEproof}
	It follows from the proof of \cref{prop:max_snr} with the caveat that, since the channel estimate $\hat{\mathbf{h}}$ is available only at the receiver, the optimal input distribution $p_{\mathbf{X}}(\bm{X})$ is uniform. This is because the channel coefficients in vector $\bar{\mathbf{h}}$ \eqref{eq:vec_ch_coeff} have uniformly distributed phases (see \cite[Sec. VII]{kramer2005cooperative}).
\end{IEEEproof}

\begin{corollary}[\!\!{\cite[Proposition 2]{karasik2020ISIT}}]\label{cor:snr_perfect}
	When the \ac{CSI} is perfectly available at both the receiver and transmitter, a transmission scheme that selects the  phase shift vector $\pmb{\theta}(\bar{\bm{h}})$ as a function of $\bar{\bm{h}}$ achieves the rate
	\begin{IEEEeqnarray}{l}\label{eq:rate_max_snr_perfect}
		R_\text{max-SNR}^\text{perfect}
		= -N\log_2(e)
		-\min_{\substack{\pmb{\theta}(\bar{\bm{h}}):\\\pmb{\theta}(\bar{\bm{h}})\in\mathcal{A}^{K\times 1}}}
		\min_{\substack{
				p_{\mathbf{X}|\bar{\mathbf{h}}}(\bm{X}|\bar{\bm{h}}):\\
				\mathbb E[\tr(\mathbf{X}\mathbf{X}^*)]\leq Km\ell,\\
				\mathbf{X}\in\mathcal C(\pmb{\theta}(\bar{\bm{h}}))^{1\times \ell}}}
			\frac{1}{m\ell}\kappa(\tilde{\mathrm{u}}|\mathbf{X}_1,\mathbf{z},\bar{\mathbf{h}}),
	\end{IEEEeqnarray}
	where the random variable $\tilde{\mathrm{u}}$ is defined as in \eqref{eq:def_tilde_u}
	with independent random vectors $\mathbf{z}\sim\mathcal{CN}(\mathbf{0},\bm{I}_{Nm\ell})$, $\bar{\mathbf{h}}\sim\mathcal{CN}(\mathbf{0},\bm{I}_{NK})$, and random matrices $\mathbf{X}_1,\mathbf{X}_2\sim p_{\mathbf{X}|\bar{\mathbf{h}}}(\bm{X}|\bar{\bm{h}})$ that are conditionally independent given $\bar{\mathbf{h}}$. Furthermore, we have the high-SNR limit $\lim_{P\rightarrow\infty}R_\text{max-SNR}^\text{perfect}=\log_2(S)$.
\end{corollary}
\begin{IEEEproof}
	\looseness=-1
	It follows from the proof of \cref{prop:max_snr} by setting $\tau=0$ and $\pmb{\Gamma}_\text{MMSE}=\mathbf{0}$, since the channel vector $\bar{\mathbf{h}}$ is known to both receiver and transmitter without requiring any training.
\end{IEEEproof}

\section{Layered Encoding}\label{sec:layered}
As discussed, achieving the capacity in \eqref{eq:capacity} requires jointly encoding the message over the phase shift vector $\pmb{\uptheta}_i(t)$ and the transmitted signal $\mathbf{s}_i(t)$, while performing optimal, i.e., maximum-likelihood joint decoding at the receiver. This may be infeasible in some communication networks.
Therefore, in this section, we propose a strategy based on layered encoding and \acf{SCD} that uses only standard separate encoding and decoding procedures, while still benefiting from the modulation of information onto the state of the RIS so as to enhance the achievable rate compared with the max-SNR scheme.

To this end, the message $w$ is split into two sub-messages, or layers, $w_1$ and $w_2$, such that $w_1$, of rate $R_1$, is encoded onto the phase shift vectors $\pmb{\uptheta}_i(t)\in\mathcal{A}^K$, whereas $w_2$, of rate $R_2$, is encoded onto the transmitted signals $\mathbf{s}_i(t)=(\mathrm{s}_{i,1}(t),\ldots,\mathrm{s}_{i,m}(t))^\intercal$, for $i=\tau+1,\ldots,\ell$ and $t\in[n/T]$. In order to enable decoding using standard \ac{SCD}, the first $\mu\geq 1$ symbols in the vectors $\mathbf{s}_i(t)$ are fixed and used as additional pilot symbols. In particular, we have
\begin{IEEEeqnarray}{c}\label{eq:scd_pilots}
	\mathrm{s}_{i,q}(t)\equiv 1,\quad i=\tau+1,\ldots,\ell,~q\in[\mu],~t\in[n/T].
\end{IEEEeqnarray}
It is worth clarifying that the pilot symbols discussed in \cref{sec:training} are employed for channel estimation, while the additional pilot symbols introduced in this section facilitate the separate decoding of the two layers, as detailed next. The pilot symbols in \eqref{eq:scd_pilots} are necessary because the channel estimation pilot symbols cannot be used for \ac{SCD} since both the transmitted symbols and \ac{RIS} reflection pattern are fixed during the channel estimation phase.

By averaging the first $\mu$ columns of the received signal matrix $\mathbf{Y}_i(t)$ in \eqref{eq:equiv_ch}, we obtain
\begin{IEEEeqnarray}{c}\label{eq:PSK_ch}
	\bar{\mathbf{y}}_i(t)\triangleq \frac{1}{\sqrt{\mu}}\sum_{q=1}^{\mu}\mathbf{y}_{i,q}(t)=\sqrt{\mu}\gamma_d\mathbf{H}(t)e^{j\pmb{\uptheta}_i(t)}+\bar{\mathbf{z}}_i(t),
\end{IEEEeqnarray}
where we have defined random vector $\bar{\mathbf{z}}_i(t)\sim\mathcal{CN}(\mathbf{0},\bm{I}_N)$.
The receiver decodes layer $w_1$ based on the received matrix $\bar{\mathbf{Y}}(t)\triangleq(\bar{\mathbf{y}}_{\tau+1}(t),\ldots,\bar{\mathbf{y}}_{\ell}(t))$, which, from \eqref{eq:PSK_ch}, can be expressed as
\begin{IEEEeqnarray}{rCl}\label{eq:psk_ch_mat}
	\bar{\mathbf{Y}}(t)&=&\gamma_d\mathbf{H}(t)\mathbf{Q}(t)+\bar{\mathbf{Z}}(t),
\end{IEEEeqnarray}
where we have defined the matrix $\bar{\mathbf{Z}}(t)\triangleq(\bar{\mathbf{z}}_{\tau+1}(t),\ldots,\bar{\mathbf{z}}_{\ell}(t))\in\mathbb C^{N\times(\ell-\tau)}$, whose elements are \ac{iid} with distribution $\mathcal{CN}(0,1)$, and the phase shift matrix
\begin{IEEEeqnarray}{c}\label{eq:def_phase_shift_mat}
	\mathbf{Q}(t)\triangleq\begin{pmatrix}
		\sqrt{\mu}e^{j\uptheta_{\tau+1,1}(t)}&\cdots&\sqrt{\mu}e^{j\uptheta_{\ell,1}(t)}\\
		\vdots&\ddots&\vdots\\
		\sqrt{\mu}e^{j\uptheta_{\tau+1,K}(t)}&\cdots&\sqrt{\mu}e^{j\uptheta_{\ell,K}(t)}
	\end{pmatrix},
\end{IEEEeqnarray}
which is selected from the set
\begin{IEEEeqnarray}{c}\label{eq:def_set_Q}
	\mathcal{Q}(\ell-\tau)\triangleq \cb{\bm{Q}\in\mathbb C^{K\times(\ell-\tau)}:Q_{k,i}=\sqrt{\mu}e^{j\theta_{i,k}},~\theta_{i,k}\in\mathcal A,~k\in[K],~i=\tau+1,\ldots,\ell}.
\end{IEEEeqnarray}
By direct inspection of \eqref{eq:psk_ch_mat}, we evince that it depends only of the \ac{RIS} phase shifts, and hence layer $w_1$ can be separately decoded.
Once layer $w_1$ is decoded, the receiver reconstructs the phase shift vectors $\pmb{\uptheta}_i(t)$, which are then used to decode layer $w_2$. This strategy achieves the rate detailed in \cref{prop:layered}.

\begin{proposition}\label{prop:layered}
	A strategy based on layered encoding and \ac{SCD} achieves the rate
	\begin{IEEEeqnarray}{c}
		R_\text{layered}(\tau,\gamma_\tau,\bm{X}_{1:\tau},\mu)=R_1(\tau,\gamma_\tau,\bm{X}_{1:\tau},\mu)+R_2(\tau,\gamma_\tau,\bm{X}_{1:\tau},\mu),
	\end{IEEEeqnarray}
	where the rate $R_1(\tau,\gamma_\tau,\bm{X}_{1:\tau},\mu)$ is defined as
	\begin{IEEEeqnarray}{c}\label{eq:layered_r1}
		R_1(\tau,\gamma_\tau,\bm{X}_{1:\tau},\mu)=-\frac{N(\ell-\tau)}{m\ell}\log_2(e)
		-\frac{1}{m\ell}\kappa(\mathrm{u}_1|\mathbf{Q}_1,\bar{\mathbf{z}},\hat{\mathbf{h}})
	\end{IEEEeqnarray}
	with random variable $\mathrm{u}_1$
	\begin{IEEEeqnarray}{c}\label{eq:def_u1}
		\mathrm{u}_1\triangleq \ln\rb{\frac{|\pmb{\Gamma}(\mathbf{Q}_1)|}{|\pmb{\Gamma}(\mathbf{Q}_2)|}}-\left\lVert\bm{V}(\mathbf{Q}_1)\bar{\mathbf{z}}+\gamma_d\rb{\mathbf{Q}_1^\otimes-\mathbf{Q}_2^\otimes}\hat{\mathbf{h}}\right\rVert^2_{\pmb{\Gamma}(\mathbf{Q}_2)}
	\end{IEEEeqnarray}
	defined by independent random vectors $\bar{\mathbf{z}}\sim\mathcal{CN}(\mathbf{0},\bm{I}_{N(\ell-\tau)})$ and $\hat{\mathbf{h}}\sim\mathcal{CN}(\mathbf{0},\bm{I}_{NK}-\mathbf{\Gamma_\text{MMSE}})$, and independent random matrices $\mathbf{Q}_1,\mathbf{Q}_2\sim p_{\mathbf{Q}}(\bm{Q})=1/A^{K(\ell-\tau)}$ for all $\bm{Q}\in\mathcal{Q}(\ell-\tau)$; and where the rate $R_2(\tau,\gamma_\tau,\bm{X}_{1:\tau},\mu)$ is defined as
	\begin{IEEEeqnarray}{c}\label{eq:layered_r2}
		R_2(\tau,\gamma_\tau,\bm{X}_{1:\tau},\mu)=-\frac{N(m-\mu)(\ell-\tau)}{m\ell}\log_2(e)
		-\frac{1}{m\ell}\kappa(\mathrm{u}_2|\check{\mathbf{X}}_1,\check{\mathbf{z}},\hat{\mathbf{h}},\pmb{\uptheta}_{\tau+1},\ldots,\pmb{\uptheta}_{\ell})
	\end{IEEEeqnarray}
	with random variable $\mathrm{u}_2$
	\begin{IEEEeqnarray}{c}\label{eq:def_u2}
		\mathrm{u}_2\triangleq \ln\rb{\frac{|\pmb{\Gamma}(\check{\mathbf{X}}_1)|}{|\pmb{\Gamma}(\check{\mathbf{X}}_2)|}}-\left\lVert\bm{V}(\check{\mathbf{X}}_1)\check{\mathbf{z}}+\gamma_d\rb{\check{\mathbf{X}}_1^\otimes-\check{\mathbf{X}}_2^\otimes}\hat{\mathbf{h}}\right\rVert^2_{\pmb{\Gamma}(\check{\mathbf{X}}_2)}
	\end{IEEEeqnarray}
	defined by independent random vectors $\check{\mathbf{z}}\sim\mathcal{CN}(\mathbf{0},\bm{I}_{N(m-\mu)(\ell-\tau)})$, $\hat{\mathbf{h}}\sim\mathcal{CN}(\mathbf{0},\bm{I}_{NK}-\mathbf{\Gamma_\text{MMSE}})$, $\pmb{\uptheta}_{\tau+1},\ldots,\pmb{\uptheta}_{\ell}\sim p_{\pmb{\uptheta}}(\pmb{\theta})=1/A^{K}$ for all $\pmb{\theta}\in\mathcal A^K$, and independent random matrices $\check{\mathbf{X}}_1,\check{\mathbf{X}}_2\sim p_{\check{\mathbf{X}}|\pmb{\uptheta}_{\tau+1},\ldots,\pmb{\uptheta}_{\ell}}(\check{\bm{X}}|\pmb{\theta}_{\tau+1},\ldots,\pmb{\theta}_{\ell})=1/S^{(m-\mu)(\ell-\tau)}$ for all $\check{\bm{X}}\in\mathcal C(\pmb{\theta}_{\tau+1},\ldots,\pmb{\theta}_{\ell};\mu)$ with
	\begin{IEEEeqnarray}{c}\label{eq:def_layered_set}
		\mathcal C(\pmb{\theta}_{\tau+1},\ldots,\pmb{\theta}_{\ell};\mu)\triangleq \cb{\check{\bm{X}}: \check{\bm{X}}=(e^{j\pmb{\theta}_{\tau+1}}\check{\mathbf{s}}_{\tau+1}^\intercal,\ldots,e^{j\pmb{\theta}_{\ell}}\check{\mathbf{s}}_{\ell}^\intercal),~\check{\mathbf{s}}_i\in\mathcal S^{(m-\mu)\times 1},~i=\tau+1,\ldots,\ell}.\IEEEeqnarraynumspace
	\end{IEEEeqnarray}
	Furthermore, for $\tau\geq K$, we obtain the high-SNR limit
	\begin{IEEEeqnarray}{c}\label{eq:layered_high_snr}
			\lim_{P\rightarrow\infty}R_\text{layered}(\tau,\gamma_\tau,\bm{X}_{1:\tau},\mu)=\frac{\ell-\tau}{m\ell}\sqb{(m-\mu)\log_2\rb{S}+K\log_2\rb{A}}.
	\end{IEEEeqnarray}
\end{proposition}
\begin{IEEEproof}
	See Appendix \ref{app:proof_layered}. 
\end{IEEEproof}

Note that the layered encoding scheme does not require \ac{CSIT} since both layers are encoded independently from the channel estimate $\hat{\mathbf{h}}$.
The rate achieved by the proposed layered strategy in the case of perfect \ac{CSI} is derived in the following corollary.

\begin{corollary}[\!\!{\cite[Proposition 4]{karasik2020ISIT}}]
	\looseness=-1
	Under the assumption that perfect \ac{CSI} is available at the receiver, a strategy based on layered encoding and \ac{SCD} achieves the rate
	\begin{IEEEeqnarray}{c}
		R_\text{layered}^\text{perfect}(\mu)=R_1^\text{perfect}(\mu)+R_2^\text{perfect}(\mu),
	\end{IEEEeqnarray}
	where the rate $R_1^\text{perfect}(\mu)$ is defined as
	\begin{IEEEeqnarray}{c}
		R_1^\text{perfect}(\mu)=-\frac{N}{m}\log_2(e)
		-\frac{1}{m\ell}\kappa(\tilde{\mathrm{u}}_1|\mathbf{Q}_1,\bar{\mathbf{z}},\bar{\mathbf{h}})
	\end{IEEEeqnarray}
	with random variable $\mathrm{u}_1$
	\begin{IEEEeqnarray}{c}
		\tilde{\mathrm{u}}_1\triangleq -\left\lVert\bar{\mathbf{z}}+\gamma_d\rb{\mathbf{Q}_1^\otimes-\mathbf{Q}_2^\otimes}\bar{\mathbf{h}}\right\rVert^2
	\end{IEEEeqnarray}
	defined by independent random vectors $\bar{\mathbf{z}}\sim\mathcal{CN}(\mathbf{0},\bm{I}_{N\ell})$ and $\bar{\mathbf{h}}\sim\mathcal{CN}(\mathbf{0},\bm{I}_{NK})$, and independent random matrices $\mathbf{Q}_1,\mathbf{Q}_2\sim p_{\mathbf{Q}}(\bm{Q})=1/A^{K\ell}$ for all $\bm{Q}\in\mathcal{Q}(\ell)$ \eqref{eq:def_set_Q};
	and where the rate $R_2^\text{perfect}(\mu)$ is defined as
	\begin{IEEEeqnarray}{c}
		R_2^\text{perfect}(\mu)=-\frac{N(m-\mu)}{m}\log_2(e)
		-\frac{1}{m\ell}\kappa(\tilde{\mathrm{u}}_2|\check{\mathbf{X}}_1,\check{\mathbf{z}},\bar{\mathbf{h}},\pmb{\uptheta}_{\tau+1},\ldots,\pmb{\uptheta}_{\ell})
	\end{IEEEeqnarray}
	with random variable $\mathrm{u}_2$
	\begin{IEEEeqnarray}{c}
		\tilde{\mathrm{u}}_2\triangleq -\left\lVert\check{\mathbf{z}}+\gamma_d\rb{\check{\mathbf{X}}_1^\otimes-\check{\mathbf{X}}_2^\otimes}\bar{\mathbf{h}}\right\rVert^2
	\end{IEEEeqnarray}
	defined by independent random vectors $\check{\mathbf{z}}\sim\mathcal{CN}(\mathbf{0},\bm{I}_{N(m-\mu)\ell})$, $\bar{\mathbf{h}}\sim\mathcal{CN}(\mathbf{0},\bm{I}_{NK})$, $\pmb{\uptheta}_{1},\ldots,\pmb{\uptheta}_{\ell}\sim p_{\pmb{\uptheta}}(\pmb{\theta})=1/A^{K}$ for all $\pmb{\theta}\in\mathcal A^K$, and independent random matrices $\check{\mathbf{X}}_1,\check{\mathbf{X}}_2\sim p_{\check{\mathbf{X}}|\pmb{\uptheta}_{1},\ldots,\pmb{\uptheta}_{\ell}}(\check{\bm{X}}|\pmb{\theta}_{1},\ldots,\pmb{\theta}_{\ell})=1/S^{(m-\mu)\ell}$ for all $\check{\bm{X}}\in\mathcal C(\pmb{\theta}_{1},\ldots,\pmb{\theta}_{\ell};\mu)$ \eqref{eq:def_layered_set}.
\end{corollary}
\begin{IEEEproof}
	\looseness=-1
	It follows from the proof of \cref{prop:layered} by setting $\tau=0$ and $\pmb{\Gamma}_\text{MMSE}=\mathbf{0}$ since the channel vector $\bar{\mathbf{h}}$ is known to both the receiver and transmitter without requiring any training.
\end{IEEEproof}

\section{Lower Bounds}\label{sec:bounds}
As discussed in the previous sections, calculating the capacity and achievable rates typically requires the evaluation of expectations over Gaussian random vectors and over discrete random matrices whose size increases exponentially with $\ell-\tau$. This makes the evaluation numerically difficult for long coherence blocks.
Furthermore, unlike the Gaussian vectors that have a known distribution, the input distribution of the random matrices needs to be numerically optimized. This implies that the standard method for estimating the expectations via empirical averages cannot be applied to the discrete random matrices, and hence estimating the expectations from a small number of samples requires methods such as the Monte Carlo gradient estimation \cite{mohamed2019monte}. In this section, we take a different approach and present lower bounds on the capacity and achievable rates that require summing over a fixed number of terms that does not increase with the number of sub-blocks $\ell$, which simplifies the exact calculation of the bounds.

\subsection{Lower Bounds for Optimal Signalling and Max-SNR}
\begin{proposition}\label{prop:capacity_lb}
	When the \ac{MMSE} estimate $\hat{\mathbf{h}}$ in \eqref{eq:mmse_ch_estimate} is available at both the receiver and transmitter, the capacity in \cref{prop:capacity} and the rate achieved by the max-SNR scheme in \cref{prop:max_snr} are lower bounded as $C(\tau,\gamma_\tau,\bm{X}_{1:\tau})\geq \underline{C}(\tau,\gamma_\tau,\bm{X}_{1:\tau})$ and $R_\text{max-SNR}(\tau,\gamma_\tau,\bm{X}_{1:\tau})\geq \underline{R}_\text{max-SNR}(\tau,\gamma_\tau,\bm{X}_{1:\tau})$, respectively, where 
	\begin{IEEEeqnarray}{l}\label{eq:lb_capacity}
		\underline{C}(\tau,\gamma_\tau,\bm{X}_{1:\tau})
		\triangleq   -\frac{N(\ell-\tau)}{\ell}\log_2(e)
		-\min_{\substack{
				p_{\mathbf{X}|\hat{\mathbf{h}}}(\bm{X}|\hat{\bm{h}}):\\
				\mathbb E[\tr(\mathbf{X}\mathbf{X}^*)]\leq Km,\\
				\mathbf{X}\in\mathcal C}}
		\frac{\ell-\tau}{m\ell}\kappa(\mathrm{u}|\mathbf{X}_1,\mathbf{z},\hat{\mathbf{h}}),
	\end{IEEEeqnarray}
	and
	\begin{IEEEeqnarray}{l}\label{eq:lb_max-snr}
		\underline{R}_\text{max-SNR}(\tau,\gamma_\tau,\bm{X}_{1:\tau})
		\triangleq 
		-\frac{N(\ell-\tau)}{\ell}\log_2(e)
		-\min_{\substack{\pmb{\theta}(\hat{\bm{h}}):\\\pmb{\theta}(\hat{\bm{h}})\in\mathcal{A}^{K\times 1}}}
		\min_{\substack{
				p_{\mathbf{X}|\hat{\mathbf{h}}}(\bm{X}|\hat{\bm{h}}):\\
				\mathbb E[\tr(\mathbf{X}\mathbf{X}^*)]\leq Km,\\
				\mathbf{X}\in\mathcal C(\pmb{\theta}(\hat{\bm{h}}))}}
		\frac{\ell-\tau}{m\ell}\kappa(\mathrm{u}|\mathbf{X}_1,\mathbf{z},\hat{\mathbf{h}}).\IEEEeqnarraynumspace
	\end{IEEEeqnarray}
	The random variable $\mathrm{u}$ in \eqref{eq:lb_capacity} and \eqref{eq:lb_max-snr} is defined as in \eqref{eq:def_u} 
	with independent random vectors $\mathbf{z}\sim\mathcal{CN}(\mathbf{0},\bm{I}_{Nm})$, $\hat{\mathbf{h}}\sim\mathcal{CN}(\mathbf{0},\bm{I}_{NK}-\mathbf{\Gamma_\text{MMSE}})$, and random matrices $\mathbf{X}_1,\mathbf{X}_2\sim p_{\mathbf{X}|\hat{\mathbf{h}}}(\bm{X}|\hat{\bm{h}})$ that are conditionally independent given $\hat{\mathbf{h}}$.
\end{proposition}
\begin{IEEEproof}
	See Appendix \ref{app_proof_cap_lb}.
\end{IEEEproof}

\looseness=-1
As detailed in Appendix \ref{app_proof_cap_lb}, the lower bounds in \cref{prop:capacity_lb} correspond to rates achievable when the sub-blocks $\bar{\mathbf{X}}_i\in\mathcal{C}$, $i=\tau+1,\ldots,\ell$, are decoded separately. This is in contrast to the optimal strategy presented in \cref{prop:capacity} that jointly decodes all data sub-blocks inputs $(\bar{\mathbf{X}}_{\tau+1},\ldots,\bar{\mathbf{X}}_\ell)\in\mathcal{C}^{\ell-\tau}$ from the channel outputs $\mathbf{y}_{\tau+1},\ldots,\mathbf{y}_\ell$. The key computational advantage of the lower bounds is that evaluating the expectations over the discrete random matrices $\mathbf{X}_1$ and $\mathbf{X}_2$ defined in \cref{prop:capacity} requires summing over $|\mathcal{C}|^{\ell-\tau}$ terms, whereas evaluating the expectations in the lower bound \eqref{eq:lb_capacity} requires summing over $|\mathcal{C}|$ terms, which is exponentially smaller.

\looseness=-1
The corresponding lower bounds on capacity and rate achieved by the max-SNR scheme under the assumptions of imperfect \ac{CSI} available only at the receiver and perfect \ac{CSI} available at both the transmitter and receiver, are formulated, respectively, in the following two corollaries.
\begin{corollary}\label{cor:lb_cap_csir}
	When the \ac{MMSE} estimate $\hat{\mathbf{h}}$ in \eqref{eq:mmse_ch_estimate} is available only at the receiver, the capacity in \cref{cor:cap_csir} and the rate achieved by the max-SNR scheme in \cref{cor:snr_csir} are lower bounded as $C_\text{CSIR}(\tau,\gamma_\tau,\bm{X}_{1:\tau})\geq \underline{C}_\text{CSIR}(\tau,\gamma_\tau,\bm{X}_{1:\tau})$ and $R_\text{max-SNR}^\text{CSIR}(\tau,\gamma_\tau,\bm{X}_{1:\tau})\geq \underline{R}_\text{max-SNR}^\text{CSIR}(\tau,\gamma_\tau,\bm{X}_{1:\tau})$, respectively, where
	\begin{IEEEeqnarray}{l}\label{eq:lb_capacity_no_csit}
		\underline{C}_\text{CSIR}(\tau,\gamma_\tau,\bm{X}_{1:\tau})
		\triangleq 
		-\frac{N(\ell-\tau)}{\ell}\log_2(e)
		-\frac{\ell-\tau}{m\ell}\kappa(\mathrm{u}|\mathbf{X}_1,\mathbf{z},\hat{\mathbf{h}})
	\end{IEEEeqnarray}
	and
	\begin{IEEEeqnarray}{l}\label{eq:lb_max-snr_no_csit}
		\underline{R}_\text{max-SNR}^\text{CSIR}(\tau,\gamma_\tau,\bm{X}_{1:\tau})
		\triangleq 
		-\frac{N(\ell-\tau)}{\ell}\log_2(e)
		-\min_{\pmb{\theta}:\pmb{\theta}\in\mathcal{A}^{K\times 1}}
		\frac{\ell-\tau}{m\ell}\kappa(\mathrm{u}|\mathbf{X}_1,\mathbf{z},\hat{\mathbf{h}}).
	\end{IEEEeqnarray}
	The random variable $\mathrm{u}$ in \eqref{eq:lb_capacity_no_csit} and \eqref{eq:lb_max-snr_no_csit} is defined as in \eqref{eq:def_u}
	with independent random vectors $\mathbf{z}\sim\mathcal{CN}(\mathbf{0},\bm{I}_{Nm})$ and $\hat{\mathbf{h}}\sim\mathcal{CN}(\mathbf{0},\bm{I}_{NK}-\mathbf{\Gamma_\text{MMSE}})$, and independent random matrices $\mathbf{X}_1,\mathbf{X}_2\sim p_{\mathbf{X}}(\bm{X})$, where $p_{\mathbf{X}}(\bm{X})=1/|\mathcal C|$ in \eqref{eq:lb_capacity_no_csit} and $p_{\mathbf{X}}(\bm{X})=1/|\mathcal C(\pmb{\theta})|$ in \eqref{eq:lb_max-snr_no_csit}.
\end{corollary}
\begin{IEEEproof}
	It follows from the proof of \cref{prop:capacity_lb} with the caveat that, since the channel estimate $\hat{\mathbf{h}}$ is available only at the receiver, the optimal input distribution $p_{\mathbf{X}}(\bm{X})$ is uniform. This is because the channel coefficients in vector $\bar{\mathbf{h}}$ \eqref{eq:vec_ch_coeff} have uniformly distributed phases (see \cite[Sec. VII]{kramer2005cooperative}).
\end{IEEEproof}

\begin{corollary}
	When perfect \ac{CSI} is available at both the receiver and transmitter, the capacity in \cref{cor:cap_perfect} and the rate achieved by the max-SNR scheme in \cref{cor:snr_perfect} are lower bounded as $C_\text{perfect}\geq \underline{C}_\text{perfect}$ and $R_\text{max-SNR}^\text{perfect}\geq \underline{R}_\text{max-SNR}^\text{perfect}$, respectively, where
	\begin{IEEEeqnarray}{l}\label{eq:lb_capacity_perfect_csi}
		\underline{C}_\text{perfect} 
		\triangleq 
		-N\log_2(e)
		-\min_{\substack{
				p_{\mathbf{X}|\bar{\mathbf{h}}}(\bm{X}|\bar{\bm{h}}):\\
				\mathbb E[\tr(\mathbf{X}\mathbf{X}^*)]\leq Km,\\
				\mathbf{X}\in\mathcal C}}
		\frac{1}{m}\kappa(\tilde{\mathrm{u}}|\mathbf{X}_1,\mathbf{z},\bar{\mathbf{h}})
	\end{IEEEeqnarray}
	and
	\begin{IEEEeqnarray}{l}\label{eq:lb_max-snr_perfect_csi}
		\underline{R}_\text{max-SNR}^\text{perfect} 
		\triangleq 
		-N\log_2(e)
		-\min_{\substack{\pmb{\theta}(\bar{\bm{h}}):\\\pmb{\theta}(\bar{\bm{h}})\in\mathcal{A}^{K\times 1}}}
		\min_{\substack{
				p_{\mathbf{X}|\bar{\mathbf{h}}}(\bm{X}|\bar{\bm{h}}):\\
				\mathbb E[\tr(\mathbf{X}\mathbf{X}^*)]\leq Km,\\
				\mathbf{X}\in\mathcal C(\pmb{\theta}(\bar{\bm{h}}))}}
		\frac{1}{m}\kappa(\tilde{\mathrm{u}}|\mathbf{X}_1,\mathbf{z},\bar{\mathbf{h}}).
	\end{IEEEeqnarray}
	The random variable $\tilde{\mathrm{u}}$ is defined as in \eqref{eq:def_tilde_u}
	with independent random vectors $\mathbf{z}\sim\mathcal{CN}(\mathbf{0},\bm{I}_{Nm})$, $\bar{\mathbf{h}}\sim\mathcal{CN}(\mathbf{0},\bm{I}_{NK})$, and random matrices $\mathbf{X}_1,\mathbf{X}_2\sim p_{\mathbf{X}|\bar{\mathbf{h}}}(\bm{X}|\bar{\bm{h}})$ that are conditionally independent given $\bar{\mathbf{h}}$.
\end{corollary}
\begin{IEEEproof}
	\looseness=-1
	It follows from the proof of \cref{prop:capacity_lb} by setting $\tau=0$ and $\pmb{\Gamma}_\text{MMSE}=\mathbf{0}$, since the channel vector $\bar{\mathbf{h}}$ is known to both the receiver and transmitter without requiring any training.
\end{IEEEproof}

\subsection{Lower Bounds for Layered Encoding}
Similar to \cref{prop:capacity_lb}, we derive a lower bound on the rate achieved by the layered-encoding scheme introduced in \cref{sec:layered}.
\begin{proposition}\label{prop:layered_lb}
	The achievable rate of the layered encoding scheme introduced in \cref{sec:layered} is lower bounded as $R_\text{layered}(\tau,\gamma_\tau,\bm{X}_{1:\tau},\mu)\geq \underline{R}_\text{layered}(\tau,\gamma_\tau,\bm{X}_{1:\tau},\mu)$ with
	\begin{IEEEeqnarray}{c}
		\underline{R}_\text{layered}(\tau,\gamma_\tau,\bm{X}_{1:\tau},\mu)\triangleq -\frac{\ell-\tau}{m\ell}\sqb{N(m+1-\mu)\log_2(e)+\kappa(\mathrm{u}_1|\mathbf{Q}_1,\bar{\mathbf{z}},\hat{\mathbf{h}})+\kappa(\mathrm{u}_2|\check{\mathbf{X}}_1,\check{\mathbf{z}},\hat{\mathbf{h}},\pmb{\uptheta})},\IEEEeqnarraynumspace
	\end{IEEEeqnarray}
	where the random variable $u_1$ is defined as in \eqref{eq:def_u1} with	independent random vectors $\bar{\mathbf{z}}\sim\mathcal{CN}(\mathbf{0},\bm{I}_{N})$ and $\hat{\mathbf{h}}\sim\mathcal{CN}(\mathbf{0},\bm{I}_{NK}-\mathbf{\Gamma_\text{MMSE}})$, and independent random matrices $\mathbf{Q}_1,\mathbf{Q}_2\sim p_{\mathbf{Q}}(\bm{Q})=1/A^{K}$ for all $\bm{Q}\in\mathcal{Q}(1)$; and where the random variable $u_2$ is defined as in \eqref{eq:def_u2} with 	independent random vectors $\check{\mathbf{z}}\sim\mathcal{CN}(\mathbf{0},\bm{I}_{N(m-\mu)})$, $\hat{\mathbf{h}}\sim\mathcal{CN}(\mathbf{0},\bm{I}_{NK}-\mathbf{\Gamma_\text{MMSE}})$, $\pmb{\uptheta}\sim p_{\pmb{\uptheta}}(\pmb{\theta})=1/A^{K}$ for all $\pmb{\theta}\in\mathcal A^K$, and independent random matrices $\check{\mathbf{X}}_1,\check{\mathbf{X}}_2\sim p_{\check{\mathbf{X}}|\pmb{\uptheta}}(\check{\bm{X}}|\pmb{\theta})=1/S^{(m-\mu)}$ for all
	$\check{\bm{X}}\in\cb{\check{\bm{X}}: \check{\bm{X}}=e^{j\pmb{\theta}}\check{\mathbf{s}}^\intercal,~\check{\mathbf{s}}\in\mathcal S^{(m-\mu)\times 1}}$.
\end{proposition}
\begin{IEEEproof}
	See Appendix \ref{app_proof_layered_lb}.
\end{IEEEproof}

The lower bound on the rate achieved by layered encoding under the assumption of perfect \ac{CSI} available at the receiver is formulated in the following corollary.
\begin{corollary}
	When perfect \ac{CSI} is available at the receiver, the achievable rate of the layered encoding scheme introduced in \cref{sec:layered} is lower bounded as $R_\text{layered}^\text{perfect}(\mu)\geq \underline{R}_\text{layered}^\text{perfect}(\mu)$ with
	\begin{IEEEeqnarray}{c}
		\underline{R}_\text{layered}^\text{perfect}(\mu)\triangleq -\frac{1}{m}\sqb{N(m+1-\mu)\log_2(e)+\kappa(\mathrm{u}_1|\mathbf{Q}_1,\bar{\mathbf{z}},\bar{\mathbf{h}})+\kappa(\mathrm{u}_2|\check{\mathbf{X}}_1,\check{\mathbf{z}},\bar{\mathbf{h}},\pmb{\uptheta})},\IEEEeqnarraynumspace
	\end{IEEEeqnarray}
	where the random variable $u_1$ is defined as in \eqref{eq:def_u1} with	independent random vectors $\bar{\mathbf{z}}\sim\mathcal{CN}(\mathbf{0},\bm{I}_{N})$ and $\bar{\mathbf{h}}\sim\mathcal{CN}(\mathbf{0},\bm{I}_{NK})$, and independent random matrices $\mathbf{Q}_1,\mathbf{Q}_2\sim p_{\mathbf{Q}}(\bm{Q})=1/A^{K}$ for all $\bm{Q}\in\mathcal{Q}(1)$; and where the random variable $u_2$ is defined as in \eqref{eq:def_u2} with independent random vectors $\check{\mathbf{z}}\sim\mathcal{CN}(\mathbf{0},\bm{I}_{N(m-\mu)})$, $\bar{\mathbf{h}}\sim\mathcal{CN}(\mathbf{0},\bm{I}_{NK})$, $\pmb{\uptheta}\sim p_{\pmb{\uptheta}}(\pmb{\theta})=1/A^{K}$ for all $\pmb{\theta}\in\mathcal A^K$, and independent random matrices $\check{\mathbf{X}}_1,\check{\mathbf{X}}_2\sim p_{\check{\mathbf{X}}|\pmb{\uptheta}}(\check{\bm{X}}|\pmb{\theta})=1/S^{(m-\mu)}$ for all
	$\check{\bm{X}}\in\cb{\check{\bm{X}}: \check{\bm{X}}=e^{j\pmb{\theta}}\check{\mathbf{s}}^\intercal,~\check{\mathbf{s}}\in\mathcal S^{(m-\mu)\times 1}}$.
\end{corollary}
\begin{IEEEproof}
	\looseness=-1
	It follows from the proof of \cref{prop:layered_lb} by setting $\tau=0$ and $\pmb{\Gamma}_\text{MMSE}=\mathbf{0}$, since the channel vector $\bar{\mathbf{h}}$ is known to both the receiver and transmitter without requiring any training.
\end{IEEEproof}

\section{Numerical Results}\label{sec:numerical}
In this section, we illustrate and discuss numerical examples with the main aims of (i) comparing the capacity achieved by the proposed joint encoding scheme with the achievable rates attained by the max-SNR and the layered encoding schemes, and (ii) assessing the impact of imperfect \ac{CSI}. For the phase response set, we consider $A$ uniformly spaced phases in the set $\mathcal A\triangleq\{0,2\pi/A,\ldots,2\pi(A-1)/A\}$, whereas, for the input constellation, we consider \ac{ASK}, which was shown to maximize capacity in the high-SNR regime (\cref{prop:capacity}), and \ac{PSK} modulations.
In addition, we set an equal power for training and data sub-blocks, i.e., $\gamma_\tau=\gamma_d=\sqrt{P}$, and optimize the channel estimation by testing all pilot symbols $\bm{X}_{1:\tau}\in\mathcal{C}^{1\times\tau}$ that satisfy the power constraint in \eqref{eq:pilots_power_constaint}. Moreover, the empirical average over Gaussian random vectors, e.g., $\hat{\mathbf{h}}$ and $\mathbf{z}$ in \cref{prop:capacity}, is evaluated via a Monte Carlo method, and the optimal input distributions, e.g., $p_{\mathbf{X}|\hat{\mathbf{h}}}(\bm{X}|\hat{\bm{h}})$ in \cref{prop:capacity}, are numerically calculated using the \emph{fmincon} function in MATLAB. We limit our investigation to small number of \ac{RIS} elements $K$ in order to perform numerical optimization without requiring excessive computing power. Based on the high-SNR analysis in \cref{prop:capacity}, we can conclude that the capacity increases linearly with the number of elements $K$ for sufficiently high \ac{SNR} and a sufficiently long coherence block. We postpone the numerical analysis with larger $K$ to future work.

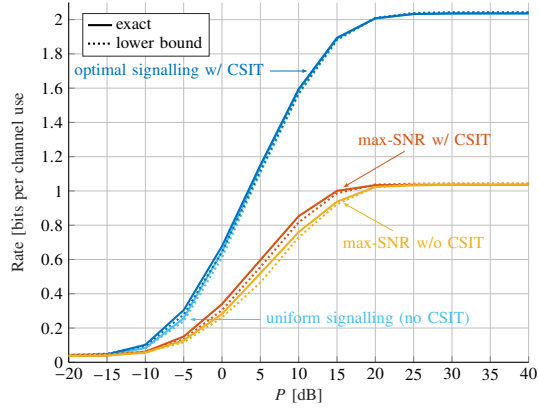
\begin{figure}[!t]
	\centering
	\resizebox {0.5\linewidth} {!} {
		\input{workspace_08-09-2020-1737.tex}
	}
	\caption{Rates as a function of the normalized power $P$ [dB] for $\ell=4$, $\tau=2$, $N=2$, $K=2$, $A=2$, $m=1$, and 4-ASK input constellation.}
	\label{fig:vs_P_good_channel_estimate}
\end{figure}

\emph{On the role of the SNR level}.
In \cref{fig:vs_P_good_channel_estimate}, we plot the rate as a function of the average power $P$, with $\ell=4$ sub-blocks of which $\tau=2$ sub-blocks are used for channel estimation, $N=2$ receive antennas, $K=2$ RIS elements, $A=2$ available phase shifts, a symbol-to-RIS control rate $m=1$, and input constellation given by the 4-ASK $\mathcal S=\{\sigma,3\sigma,5\sigma,7\sigma\}$ with $\sigma=1/\sqrt{21}$.
For very low SNR, i.e., less than $-20$dB, it is observed that the max-SNR approach is close to being optimal, and hence, in this regime, encoding information in the RIS reflection pattern does not increase the rate. For larger SNR levels of practical interest, however, joint encoding provides significant gain over the max-SNR scheme. 

It is also observed that \ac{CSIT} is unnecessary for very low or very high \ac{SNR} levels. This is because, at low \ac{SNR}, the channel estimate is poor and cannot be applied for beamforming, whereas, at high \ac{SNR}, beamforming, which is used to increase \ac{SNR}, is unnecessary.
In addition, the lower bounds presented in \cref{sec:bounds} are shown to be close to the achievable rates. Note that the gap to the lower bounds increases for small number of pilot symbols $\tau<K$, i.e., when channel estimation is poor, even for high-SNR.

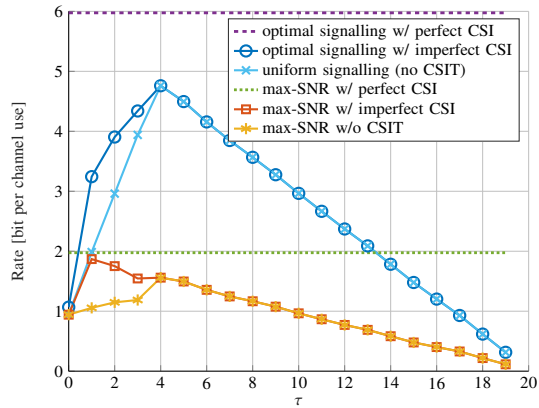
\begin{figure}[!t]
	\centering
	\resizebox {0.5\linewidth} {!} {	
		\input{workspace_08-11-2020-1011.tex}
	}
	\caption{Rate lower bounds as a function of the number of training sub-blocks $\tau$ for $\ell=20$, $N=2$, $K=4$, $A=2$, $m=1$, $P=40$ dB, and 4-ASK input constellation.}
	\label{fig:vs_tau}
\end{figure}

\emph{Optimal number of pilot symbols.}
In \cref{fig:vs_tau}, we plot the lower bounds on the rate as a function of the number of training sub-blocks $\tau$ with $\ell=20$ sub-blocks in each coherence block, $N=2$ receive antennas, $K=4$ RIS elements, $A=2$ available phase shifts, a symbol-to-RIS control rate $m=1$, an average power constraint of $P=40$ dB, and an input constellation given by 4-ASK. 
\looseness=-1
Note that we plot the lower bounds and not exact expressions since evaluating the capacity requires summing over the set of channel inputs $\mathbf{X}$ whose size is $|\mathcal C|^{\ell-\tau}=(A^K\cdot S^m)^{\ell-\tau}=2^{120-6\tau}$, which is not feasible.
It is observed that the lower bound on the capacity increases with $\tau$ up to $\tau=4$, and then decreases. This is because increasing the number of pilot symbols improves channel estimation accuracy on the one hand, but on the other hand leaves fewer sub-blocks for transmitting data. 
In addition, joint encoding is shown to require a more accurate channel estimation compared to the max-SNR scheme with \ac{CSIT}, for which allocating $\tau=1$ pilot is optimal.
Comparing the penalty of channel estimation between the joint encoding strategy and the max-SNR scheme, in addition, we observe that the gap is larger for joint encoding since a higher percentage of the coherence block is used to obtain a sufficient channel estimation accuracy.

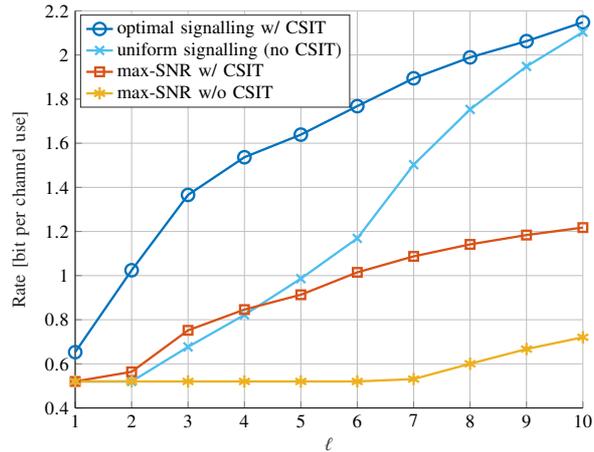
\begin{figure}[!t]
	\centering
	\resizebox {0.5\linewidth} {!} {	
		\input{workspace_08-12-2020-1035.tex}
	}
	\caption{Rate lower bounds as a function of the number of sub-blocks $\ell$ for $N=1$, $K=4$, $A=2$, $m=1$, $P=10$ dB, and 4-ASK input constellation.}
	\label{fig:vs_ell}
\end{figure}

As seen in \cref{fig:vs_tau}, the capacity-achieving joint encoding strategy requires a better channel estimation compared to the max-SNR scheme. However, for short coherence blocks, acquiring sufficiently good channel estimation might not be feasible and the gain of joint encoding is expected to decrease. This is illustrated in \cref{fig:vs_ell}, where we plot the lower bounds on the rate as a function of the number of sub-blocks $\ell$ with $N=2$ receive antennas, $K=4$ RIS elements, $A=2$ available phase shifts, a symbol-to-RIS control rate $m=1$, an average power constraint of $P=10$ dB, and an input constellation given by 4-ASK. For each value of $\ell$, the lower bounds are optimized over $\tau=0,\ldots,\ell-1$.
\looseness=-1
For fast-changing channels, the gain of joint encoding is shown to be low. Moreover, without \ac{CSIT}, the max-SNR scheme is optimal for $\ell\leq 2$.

\begin{figure}[!t]
	\centering
	\resizebox {0.5\linewidth} {!} {	
		\input{workspace_08-12-2020-1417.tex}
	}
	\caption{Rate lower bounds as a function of the number of receive antennas $N$ for $\ell=30$, $\tau=6$, $K=6$, $A=2$, $m=1$, $P=10$ dB, and 2-ASK input constellation.}
	\label{fig:vs_N}
\end{figure}
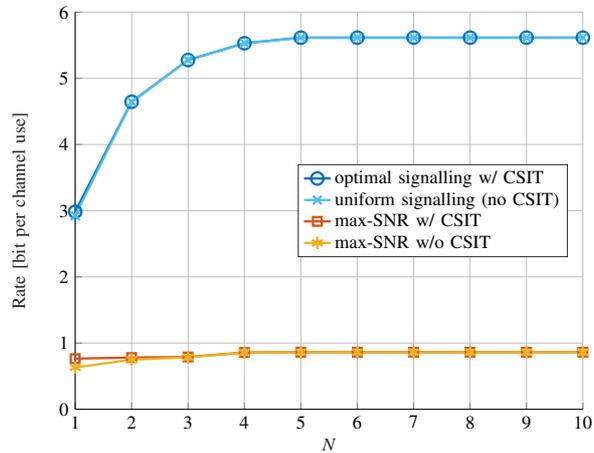

\emph{On the number of receive antennas.}
In \cref{fig:vs_N}, we plot the lower bounds on the rate as a function of the number of receive antennas $N$ with $\ell=30$ sub-blocks of which $\tau=6$ sub-blocks are used for channel estimation, $K=6$ RIS elements, $A=2$ available phase shifts, a symbol-to-RIS control rate $m=1$, an average power constraint of $P=10$ dB, and an input constellation given by 2-ASK $\mathcal S=\{\sigma,3\sigma\}$ with $\sigma=1/\sqrt{5}$. 
While both capacity and rate achieved by the max-SNR scheme increase with the number of receive antennas, the effect is more prominent for joint encoding since, for the max-SNR scheme, spatial multiplexing is restricted by the number of transmit antennas, whereas, for joint encoding, spatial multiplexing is restricted by the number of \ac{RIS} elements.

\begin{figure}[!t]
	\centering
	\resizebox {0.5\linewidth} {!} {
		\input{workspace_09-11-2020-1314.tex}
	}
	\caption{Rate lower bounds as a function of the normalized power $P$ [dB] for $\ell=50$, $\tau=3$, $N=2$, $K=3$, $A=2$, $m=2$, $\mu=1$, and 4-ASK or QPSK input constellation.}
	\label{fig:layered}
\end{figure}
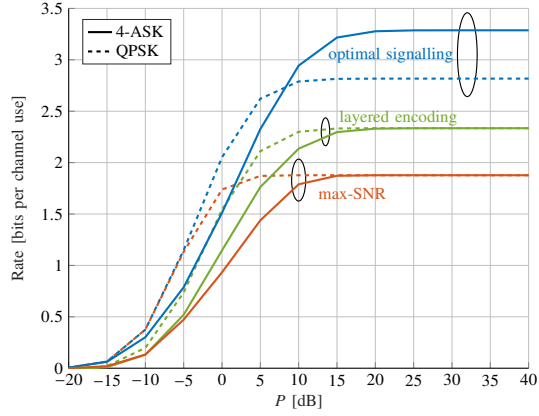

\emph{Layered Encoding.}
In \cref{fig:layered}, we compare the rate achieved by layered encoding to that of the max-SNR method and to the capacity by plotting the lower bounds on the rate as a function of the average power $P$, with $\ell=50$ sub-blocks of which $\tau=3$ sub-blocks are used for channel estimation, $N=2$ receive antennas, $K=3$ RIS elements, $A=2$ available phase shifts, a symbol-to-RIS control rate $m=2$, and input constellation given by 4-ASK or QPSK $\mathcal S=\{\pm 1,\pm i\}$. For layered encoding, we set $\mu=1$ pilot, which was seen to maximize the rate in this experiment. 
It is observed that, for sufficiently high SNR, the layered-encoding scheme improves over the max-SNR approach. Note that, in the high-SNR regime, as apparent from the limits in \eqref{eq:max-snr_high_snr} and \eqref{eq:layered_high_snr}, layered encoding achieves a higher rate when $K\log_2(A)>\mu\log_2(S)$.
In addition, while \ac{PSK} outperforms \ac{ASK} when used with the max-SNR and layered-encoding schemes, the opposite is true with joint encoding in the high-SNR regime. In fact, as discussed in \cref{prop:capacity}, in the high-SNR regime, out of all finite input sets $\mathcal S$ with the same size, \ac{ASK} achieves the maximum capacity.

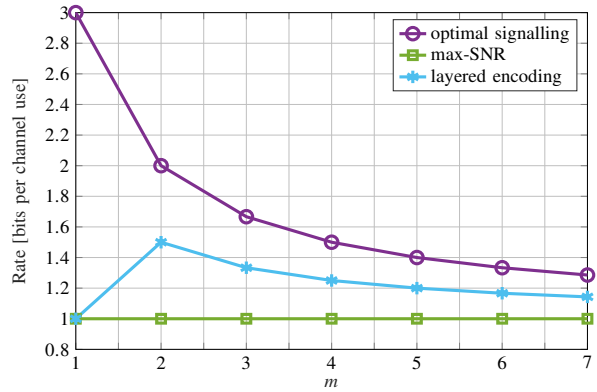
\begin{figure}[!t]
	\centering
	\resizebox {0.5\linewidth} {!} {	
		\input{N=2_K=2_A=2_P=40_numInputs=2_vs_DT.tex}
	}
	\caption{Rates with perfect \ac{CSI} as a function of the RIS control rate factor $m$ for $N=2$, $K=2$, $A=2$, $P=40$ dB, $\mu=1$, and 2-ASK input constellation.}
	\label{fig:cap_vs_m_2ASK_A=2_P=40}
\end{figure}

\emph{On the \ac{RIS} control rate.}
The gain of using the state of the RIS as a medium for conveying information is expected to decrease as the rate of the control link from the transmitter to the RIS decreases. This is illustrated in \cref{fig:cap_vs_m_2ASK_A=2_P=40}, where we plot the rate with perfect \ac{CSI} at both transmitter and receiver as a function of the RIS control rate factor $m$, with $N=2$ receive antennas, $K=2$ RIS elements, $A=2$ available phase shifts, an average power constraint of $P=40$ dB, and an input constellation 2-ASK.
Note that the performance of the layered-encoding scheme improves from $m=1$ to $m=2$ since, for $m=1$, the transmitted symbol in each sub-block is used as a pilot, and hence only the first layer carries information.
It is observed that, while, for $m=1$, joint encoding achieves three times the rate of max-SNR, the gain reduces to a factor of $1.3$ for $m=7$.

\section{Conclusions}\label{sec:conclusions}
In this work, we have studied the capacity of an \ac{RIS}-aided system. We focused on a fundamental model with one transmitter and one receiver, where the \ac{CSI} is acquired through pilot-assisted channel estimation. The common approach of using the \ac{RIS} as a passive beamformer to maximize the achievable rate was shown to be generally suboptimal in terms of the achievable rate for finite input constellations, especially for slow-changing channels. Instead, the capacity-achieving scheme was proved to jointly encode information in the RIS reflection pattern as well as in the transmitted signal. While the scheme was shown to require a more accurate channel estimation compared to the max-SNR approach, the gain of encoding information in the reflection pattern of the \ac{RIS} was demonstrated to be significant for a sufficiently high \ac{RIS} control rate. In addition, a suboptimal, yet practical, strategy based on separate layered encoding and successive cancellation decoding was demonstrated to outperform passive beamforming for sufficiently high SNR levels, and motivates \ac{RIS}-based modulation design \cite{basar2019Media,yan2020passive,lin2020reconfigurable,basar2020Reconfigurable,li2020single,tang2020Wireless} for single-\ac{RF} MIMO communication.

Among related problems left open by this study, we mention the design of low-complexity joint encoding and decoding strategies that approach capacity, the derivation of the capacity for noisy RIS \cite{qian2020beamforming} and for \ac{RIS} with mutual coupling \cite{gradoni2020end}, and extensions to RIS systems with multiple users/surfaces
\cite{guo2019weighted}
or with security constraints \cite{guan2019intelligent}.  
Another related problem is finding the optimal input distribution for a slowly fading channel with \ac{CSI} only at the receiver \cite{shamai2003broadcast}. 

\appendix

\subsection{Proof of Proposition \ref{prop:capacity}}\label{app_proof_cap}
The model in \eqref{eq:vec_data_ch} can be viewed as a standard channel with input $\mathbf{X}$, output $\mathbf{y}$, and known \ac{CSI} $\hat{\mathbf{h}}$. This is because the transmitter directly controls the states of the RIS $\pmb{\uptheta}_i(t)$ and the transmitted symbols $\mathbf{s}_i(t)$ for $i\in[\ell]$ and $t\in[n/m]$. Therefore, it follows from the channel coding theorem \cite[Ch. 7]{cover2006elements}, \cite[Ch. 7.4.1]{el2011network}, that the ergodic capacity can be expressed as
\begin{IEEEeqnarray}{rCl}\label{eq:capacity_implicit}
	C(\tau,\gamma_\tau,\bm{X}_{1:\tau})
	&=&\max_{\substack{
			p_{\mathbf{X}|\hat{\mathbf{h}}}(\bm{X}|\hat{\bm{h}}):\\
			\mathbb E[\tr(\mathbf{X}\mathbf{X}^*)]\leq Km(\ell-\tau),\\
			\mathbf{X}\in\mathcal C^{1\times (\ell-\tau)}}}
		\frac{1}{m\ell}I(\mathbf{X};\mathbf{y}|\hat{\mathbf{h}}).
\end{IEEEeqnarray}
The mutual information $I(\mathbf{X};\mathbf{y}|\hat{\mathbf{h}})$ in \eqref{eq:capacity_implicit} can be written as $I(\mathbf{X};\mathbf{y}|\hat{\mathbf{h}})=h(\mathbf{y}|\hat{\mathbf{h}})-h(\mathbf{y}|\hat{\mathbf{h}},\mathbf{X}).$
In addition, the conditional probability density function of the output $\mathbf{y}$ given the estimate $\hat{\mathbf{h}}$ and input $\mathbf{X}$ is
\begin{IEEEeqnarray}{c}
	p_{\mathbf{y}|\hat{\mathbf{h}},\mathbf{X}}(\bm{y}|\hat{\bm{h}},\bm{X})=\frac{1}{\pi^{Nm(\ell-\tau)}|\pmb{\Gamma}(\bm{X})|}\exp\cb{-\left\lVert\bm{y}-\gamma_d\bm{X}^\otimes\hat{\bm{h}}\right\rVert^2_{\pmb{\Gamma}(\bm{X})}},
\end{IEEEeqnarray}
where the covariance matrix $\pmb{\Gamma}(\bm{X})$ is defined in \eqref{eq:gamma_X_def}.
Therefore, the conditional differential entropy $h(\mathbf{y}|\hat{\mathbf{h}},\mathbf{X})$ is given as
\begin{IEEEeqnarray}{c}\label{eq:app_cond_diff_ent1}
	h(\mathbf{y}|\hat{\mathbf{h}},\mathbf{X})=Nm(\ell-\tau)\log_2(\pi e)+\int_{\mathbb C^{NK\times 1}}p_{\hat{\mathbf{h}}}(\hat{\bm{h}})\sum_{\bm{X}\in\mathcal C^{1\times(\ell-\tau)}}p_{\mathbf{X}|\hat{\mathbf{h}}}(\bm{X}|\hat{\bm{h}})\log_2\det\rb{\pmb{\Gamma}(\bm{X})}\diff{\hat{\bm{h}}},\IEEEeqnarraynumspace
\end{IEEEeqnarray}
and the conditional differential entropy $h(\mathbf{y}|\hat{\mathbf{h}})$ can be expressed as (see, e.g., \cite[Eq. (3)]{ungerboeck1982Channel} and \cite[Eq. (4)]{he2005Computing})
\begin{IEEEeqnarray}{rCl}\label{eq:app_cond_diff_ent2}
	h(\mathbf{y}|\hat{\mathbf{h}})
	&=&Nm(\ell-\tau)\log_2(\pi)\\
	&&-\int_{\mathbb C^{NK\times 1}}\!\!\!\! p_{\hat{\mathbf{h}}}(\hat{\bm{h}})\int_{\mathbb C^{Nm(\ell-\tau)\times 1}}\!\!\!\!\!\! p_{\mathbf{z}}(\bm{z})\sum_{\bm{X}_1\in\mathcal C^{1\times(\ell-\tau)}}\!\!\!\! p_{\mathbf{X}|\hat{\mathbf{h}}}(\bm{X}_1|\hat{\bm{h}})\log_2\rb{\sum_{\bm{X}_2\in\mathcal C^{1\times(\ell-\tau)}}\!\!\!\! \frac{p_{\mathbf{X}|\hat{\mathbf{h}}}(\bm{X}_2|\hat{\bm{h}})}{|\pmb{\Gamma}(\bm{X}_1)|}e^u}\diff{\bm{z}}\diff{\hat{\bm{h}}}\IEEEeqnarraynumspace\IEEEnonumber
\end{IEEEeqnarray}
with $\mathbf{z}\sim\mathcal{CN}(\mathbf{0},\bm{I}_{Nm(\ell-\tau)})$ and where we have defined the scalar
\begin{IEEEeqnarray}{c}
	u\triangleq \ln\rb{\frac{|\pmb{\Gamma}(\bm{X}_1)|}{|\pmb{\Gamma}(\bm{X}_2)|}}-\left\lVert\bm{V}(\bm{X}_1)\bm{z}+\gamma_d\rb{\bm{X}_1^\otimes-\bm{X}_2^\otimes}\hat{\bm{h}}\right\rVert^2_{\pmb{\Gamma}(\bm{X}_2)}.
\end{IEEEeqnarray} 
Overall, by subtracting \eqref{eq:app_cond_diff_ent1} from \eqref{eq:app_cond_diff_ent2} and applying the conditional \ac{CGF} definition in \eqref{eq:def_cond_cgf_rand}, we get \eqref{eq:capacity}.
Note that the mutual information $I(\mathbf{X};\mathbf{y}|\hat{\mathbf{h}})$ is a concave function of $p_{\mathbf{X}|\hat{\mathbf{h}}}(\bm{X}|\hat{\bm{h}})$ for fixed $p_{\mathbf{y}|\hat{\mathbf{h}},\mathbf{X}}(\bm{y}|\hat{\bm{h}},\bm{X})$ \cite[Theorem 2.7.4]{cover2006elements}, Therefore, problem~\eqref{eq:capacity_implicit} can be solved using convex optimization tools.

In the high-SNR regime, the channel can be perfectly estimated if $\tau\geq K$, and hence we have the limit
\begin{IEEEeqnarray}{c}
	\lim_{P\rightarrow\infty}I(\mathbf{X};\mathbf{y}|\hat{\mathbf{h}}) = H(\mathbf{X}|\hat{\mathbf{h}})\leq H(\mathbf{X})\leq (\ell-\tau)\log_2(|\mathcal{C}|),
\end{IEEEeqnarray}
where equality is achieved for a uniform distribution  $p_{\mathbf{X}|\hat{\mathbf{h}}}(\bm{X}|\hat{\bm{h}})=1/|\mathcal{C}|^{\ell-\tau}$. Furthermore, the cardinality of the set $\mathcal{C}$ in \eqref{eq:set_C} is upper bounded as $|\mathcal{C}|\leq S^m\cdot A^K$, where equality is achieved for the \ac{ASK} input constellation.

\subsection{Proof of Proposition \ref{prop:layered}}\label{app:proof_layered}
The channel in \eqref{eq:psk_ch_mat} is equivalent to a point-to-point Gaussian \ac{MIMO} channel with \ac{PSK} input $\mathbf{Q}$. Therefore, for layer $w_1$, the following rate is achievable
\begin{IEEEeqnarray}{rCl}\label{eq:app_MI_layered1}
	R_1(\tau,\gamma_\tau,\bm{X}_{1:\tau},\mu)&=& \frac{1}{m\ell}I(\mathbf{Q};\bar{\mathbf{Y}}|\hat{\mathbf{H}})
	=\frac{1}{m\ell}I(\mathbf{Q};\bar{\mathbf{y}}|\hat{\mathbf{h}}),
\end{IEEEeqnarray}
where we have defined 
\begin{IEEEeqnarray}{c}\label{eq:app_bar_y_def}
	\bar{\mathbf{y}}\triangleq\stack{(\bar{\mathbf{Y}})}=\gamma_d\mathbf{Q}^\otimes\bar{\mathbf{h}}+\bar{\mathbf{z}}
\end{IEEEeqnarray}
with $\bar{\mathbf{z}}\sim\mathcal{CN}(\bm{0},\bm{I}_{N(\ell-\tau)})$, and
where the phase shifts matrix $\mathbf{Q}$ \eqref{eq:def_phase_shift_mat} is uniformly distributed, i.e., $p_{\mathbf{Q}}(\bm{Q})=1/A^{K(\ell-\tau)}$ for all $\bm{Q}\in\mathcal{Q}(\ell-\tau)$ \eqref{eq:def_set_Q}.
It hence follows from the proof of \cref{prop:capacity} (Appendix \ref{app_proof_cap}) that we have
\begin{IEEEeqnarray}{rCl}\label{eq:app_mi_phase}
	I(\mathbf{Q};\bar{\mathbf{y}}|\hat{\mathbf{h}})&=&-N(\ell-\tau)\log_2(e)\\
	&&-\int_{\mathbb C^{NK\times 1}}p_{\hat{\mathbf{h}}}(\hat{\bm{h}})\int_{\mathbb C^{N(\ell-\tau)\times 1}}p_{\bar{\mathbf{z}}}(\bar{\bm{z}})\sum_{\bm{Q}_1\in\mathcal{Q}(\ell-\tau)}\frac{1}{A^{K(\ell-\tau)}}\log_2\rb{\sum_{\bm{Q}_2\in\mathcal{Q}(\ell-\tau)}\frac{\exp\{u_1\}}{A^{K(\ell-\tau)}}}\diff{\bar{\bm{z}}}\diff{\hat{\bm{h}}},\IEEEnonumber
\end{IEEEeqnarray}
where we have defined the scalar
\begin{IEEEeqnarray}{c}
	u_1\triangleq \ln\rb{\frac{|\pmb{\Gamma}(\bm{Q}_1)|}{|\pmb{\Gamma}(\bm{Q}_2)|}}-\left\lVert\bm{V}(\bm{Q}_1)\bar{\bm{z}}+\gamma_d\rb{\bm{Q}_1^\otimes-\bm{Q}_2^\otimes}\hat{\bm{h}}\right\rVert^2_{\pmb{\Gamma}(\bm{Q}_2)}.
\end{IEEEeqnarray}
By applying the conditional \ac{CGF} definition in \eqref{eq:def_cond_cgf_rand} to the achievable rate in \eqref{eq:app_MI_layered1} with the aid of \eqref{eq:app_mi_phase}, we get \eqref{eq:layered_r1}.

For layer $w_2$, let $\check{\mathbf{Y}}_i(t)$ denote the last $(m-\mu)$ column of $\mathbf{Y}_i(t)$ \eqref{eq:equiv_ch}, i.e.,
\begin{IEEEeqnarray}{c}
	\check{\mathbf{Y}}_i(t)\triangleq (\mathbf{y}_{i,\mu+1}(t),\ldots,\mathbf{y}_{i,m}(t))=\gamma_d\mathbf{H}(t)e^{j\pmb{\uptheta}_i(t)}\check{\mathbf{s}}_i^\intercal(t)+\check{\mathbf{Z}}_i(t),
\end{IEEEeqnarray}
where we have defined $\check{\mathbf{s}}_i(t)\triangleq (\mathrm{s}_{i,\mu+1},\ldots,\mathrm{s}_{i,m})^\intercal\in\mathcal S^{(m-\mu)\times 1}$ and $\check{\mathbf{Z}}_i(t)\in\mathbb C^{N\times(m-\mu)}$ whose elements are \ac{iid} as $\mathcal{CN}(0,1)$. After layer $w_1$ is decoded, the receiver reconstructs the phase shifts $\{\pmb{\theta}_i(t)\}_i$, $i=\tau+1,\ldots,\ell$, $t\in[n/T]$, and decodes layer $w_2$ from the received signals
\begin{IEEEeqnarray}{c}\label{eq:app_check_Y}
	\check{\mathbf{Y}}(t)\triangleq (\check{\mathbf{Y}}_{\tau+1}(t),\ldots,\check{\mathbf{Y}}_{\ell}(t))=\gamma_d\mathbf{H}(t)\check{\mathbf{X}}(t)+\check{\mathbf{Z}}(t),
\end{IEEEeqnarray}
where we have defined $\check{\mathbf{Z}}(t)\triangleq (\check{\mathbf{Z}}_{\tau+1}(t),\ldots,\check{\mathbf{Z}}_{\ell}(t))$ and
\begin{IEEEeqnarray}{c}\label{eq:app_def_check_X}
	\check{\mathbf{X}}(t)\triangleq (e^{j\pmb{\theta}_{\tau+1}(t)}\check{\mathbf{s}}_{\tau+1}^\intercal(t),\ldots,e^{j\pmb{\theta}_{\ell}(t)}\check{\mathbf{s}}_{\ell}^\intercal(t))\in \mathcal{C}(\pmb{\Theta};\mu)
\end{IEEEeqnarray}
with $\pmb{\Theta}\triangleq (\pmb{\theta}_{\tau+1},\ldots,\pmb{\theta}_{\tau+1})$.
Therefore, the following rate is achievable for layer $w_2$
\begin{IEEEeqnarray}{rCl}\label{eq:app_MI_layered2}
	R_2(\tau,\gamma_\tau,\bm{X}_{1:\tau},\mu)&=& \frac{1}{m\ell}I(\check{\mathbf{X}};\check{\mathbf{Y}}|\hat{\mathbf{h}},\pmb{\uptheta}_{\tau+1},\ldots,\pmb{\uptheta}_{\ell})
	=\frac{1}{m\ell}I(\check{\mathbf{X}};\check{\mathbf{y}}|\hat{\mathbf{h}},\pmb{\uptheta}_{\tau+1},\ldots,\pmb{\uptheta}_{\ell}),
\end{IEEEeqnarray}
where we have defined 
\begin{IEEEeqnarray}{c}\label{eq:app_def_check_y}
	\check{\mathbf{y}}\triangleq\stack{(\check{\mathbf{Y}})}=\gamma_d\check{\mathbf{X}}^\otimes\bar{\mathbf{h}}+\check{\mathbf{z}}
\end{IEEEeqnarray}
with $\check{\mathbf{z}}\sim\mathcal{CN}(\bm{0},\bm{I}_{N(m-\mu)(\ell-\tau)})$, and where the input $\check{\mathbf{X}}$ is uniformly distributed, i.e., $p_{\check{\mathbf{X}}}(\check{\bm{X}})=1/S^{(m-\mu)(\ell-\tau)}$ for all $\check{\bm{X}}\in \mathcal{C}(\pmb{\Theta};\mu)$ \eqref{eq:def_layered_set}. Similar to layer $w_1$, we have
\begin{IEEEeqnarray}{l}\label{eq:app_mi_layer2}
	I(\check{\mathbf{X}};\check{\mathbf{y}}|\hat{\mathbf{h}},\pmb{\uptheta}_{\tau+1},\ldots,\pmb{\uptheta}_{\ell})\\=-N(m-\mu)(\ell-\tau)\log_2(e)\IEEEnonumber\\
	\quad-\int_{\mathbb C^{NK\times 1}}\!\!\!\! p_{\hat{\mathbf{h}}}(\hat{\bm{h}})\int_{\mathbb C^{N(\ell-\tau)\times 1}}\!\!\!\!\!\!\!\!\!\!\!\! p_{\check{\mathbf{z}}}(\check{\bm{z}})\Bigg[\sum_{\pmb{\Theta}\in\mathcal A^{K\times(\ell-\tau)}}\!\! \frac{1}{A^{K(\ell-\tau)}}
	\sum_{\check{\bm{X}}_1\in\mathcal \mathcal{C}(\pmb{\Theta})}\!\!\! \frac{1}{S^{(m-\mu)(\ell-\tau)}}\log_2\rb{\sum_{\check{\bm{X}}_2\in\mathcal \mathcal{C}(\pmb{\Theta})}\!\!\! \frac{\exp\{u_2\}}{S^{(m-\mu)(\ell-\tau)}}}\Bigg]\diff{\check{\bm{z}}}\diff{\hat{\bm{h}}},\IEEEnonumber
\end{IEEEeqnarray}
where we have defined the scalar
\begin{IEEEeqnarray}{c}
	u_2\triangleq \ln\rb{\frac{|\pmb{\Gamma}(\check{\bm{X}}_1)|}{|\pmb{\Gamma}(\check{\bm{X}}_2)|}}-\left\lVert\bm{V}(\check{\bm{X}}_1)\check{\bm{z}}+\gamma_d\rb{\check{\bm{X}}_1^\otimes-\check{\bm{X}}_2^\otimes}\hat{\bm{h}}\right\rVert^2_{\pmb{\Gamma}(\check{\bm{X}}_2)}.
\end{IEEEeqnarray}
By applying the conditional \ac{CGF} definition in \eqref{eq:def_cond_cgf_rand} to the achievable rate in \eqref{eq:app_MI_layered2} with the aid of \eqref{eq:app_mi_layer2}, we get \eqref{eq:layered_r2}.

As in the proof of \cref{prop:capacity}, for $\tau\geq K$, we have the high-\ac{SNR} limits
\begin{IEEEeqnarray}{c}
	\lim_{P\rightarrow\infty}I(\mathbf{Q};\bar{\mathbf{y}}|\hat{\mathbf{h}}) = H(\mathbf{Q})= (\ell-\tau)K\log_2(A)
\end{IEEEeqnarray}
and
\begin{IEEEeqnarray}{c}
	\lim_{P\rightarrow\infty}I(\check{\mathbf{X}};\check{\mathbf{y}}|\hat{\mathbf{h}},\pmb{\uptheta}_{\tau+1},\ldots,\pmb{\uptheta}_{\ell}) = H(\check{\mathbf{X}}|\pmb{\uptheta}_{\tau+1},\ldots,\pmb{\uptheta}_{\ell})= (\ell-\tau)(m-\mu)\log_2(S).
\end{IEEEeqnarray}

\subsection{Proof of Proposition \ref{prop:capacity_lb}}\label{app_proof_cap_lb}
The mutual information $I(\mathbf{X};\mathbf{y}|\hat{\mathbf{h}})$ in \eqref{eq:capacity_implicit} can be lower bounded as
\begin{IEEEeqnarray}{rCl}\label{eq:MI_LB}
	I(\mathbf{X};\mathbf{y}|\hat{\mathbf{h}})&=&H(\bar{\mathbf{X}}_{\tau+1},\ldots,\bar{\mathbf{X}}_\ell|\hat{\mathbf{h}})-H(\bar{\mathbf{X}}_{\tau+1},\ldots,\bar{\mathbf{X}}_\ell|\hat{\mathbf{h}},\mathbf{y}_{\tau+1},\ldots,\mathbf{y}_\ell)\IEEEnonumber\\
	&\overset{\text{(a)}}{=}&\sum_{i=\tau+1}^{\ell}\sqb{H(\bar{\mathbf{X}}_i|\hat{\mathbf{h}})-H(\bar{\mathbf{X}}_{i}|\hat{\mathbf{h}},\mathbf{y}_{\tau+1},\ldots,\mathbf{y}_\ell,\bar{\mathbf{X}}_{\tau+1},\ldots,\bar{\mathbf{X}}_{i-1})}\IEEEnonumber\\
	&\overset{\text{(b)}}{\geq}&\sum_{i=\tau+1}^{\ell}\sqb{H(\bar{\mathbf{X}}_i|\hat{\mathbf{h}})-H(\bar{\mathbf{X}}_{i}|\hat{\mathbf{h}},\mathbf{y}_{i})}\IEEEnonumber\\
	&=&(\ell-\tau)I(\bar{\mathbf{X}}_{\tau+1};\mathbf{y}_{\tau+1}|\hat{\mathbf{h}}),
\end{IEEEeqnarray}
where the equality (a) follows from the entropy chain rule \cite[Thm. 2.5.1]{cover2006elements} and since, given the channel estimate $\hat{\mathbf{h}}$, the inputs at different sub-blocks are independent; and inequality (b) is due to the \emph{conditioning reduces entropy} property \cite[Thm. 2.6.5]{cover2006elements}. 
\cref{prop:capacity_lb} is then proved by repeating the proof in Appendix \ref{app_proof_cap} with the caveat that the mutual information $I(\mathbf{X};\mathbf{y}|\hat{\mathbf{h}})$ is replaced with the lower bound \eqref{eq:MI_LB}.

\subsection{Proof of Proposition \ref{prop:layered_lb}}\label{app_proof_layered_lb}
Similar to the proof of \cref{prop:capacity_lb} (Appendix \ref{app_proof_cap_lb}), the mutual information $I(\mathbf{Q};\bar{\mathbf{y}}|\hat{\mathbf{h}})$ in \eqref{eq:app_MI_layered1} can be lower bounded as
\begin{IEEEeqnarray}{c}\label{eq:app_layer1_mi_lb}
	I(\mathbf{Q};\bar{\mathbf{y}}|\hat{\mathbf{h}})\overset{\text{(a)}}{=}I(\pmb{\uptheta}_{\tau+1},\ldots,\pmb{\uptheta}_\ell;\bar{\mathbf{y}}_{\tau+1},\ldots,\bar{\mathbf{y}}_{\ell}|\hat{\mathbf{h}})
	\geq (\ell-\tau)I(\pmb{\uptheta}_{\tau+1};\bar{\mathbf{y}}_{\tau+1}|\hat{\mathbf{h}}),
\end{IEEEeqnarray}
where the equality (a) follows from the definitions in \eqref{eq:PSK_ch}, \eqref{eq:def_phase_shift_mat}, and \eqref{eq:app_bar_y_def}. Furthermore, the mutual information $I(\check{\mathbf{X}};\check{\mathbf{y}}|\hat{\mathbf{h}},\pmb{\uptheta}_{\tau+1},\ldots,\pmb{\uptheta}_{\ell})$ in \eqref{eq:app_MI_layered2} can be lower bounded as
\begin{IEEEeqnarray}{rCl}\label{eq:app_layer2_mi_lb}
	I(\check{\mathbf{X}};\check{\mathbf{y}}|\hat{\mathbf{h}},\pmb{\uptheta}_{\tau+1},\ldots,\pmb{\uptheta}_{\ell})&\overset{\text{(a)}}{=}&I(\check{\mathbf{s}}_{\tau+1},\ldots,\check{\mathbf{s}}_{\ell};\check{\mathbf{Y}}_{\tau+1},\ldots,\check{\mathbf{Y}}_\ell|\hat{\mathbf{h}},\pmb{\uptheta}_{\tau+1},\ldots,\pmb{\uptheta}_{\ell})\IEEEnonumber\\
	&\geq& (\ell-\tau)I(\check{\mathbf{s}}_{\tau+1};\check{\mathbf{Y}}_{\tau+1}|\hat{\mathbf{h}},\pmb{\uptheta}_{\tau+1}),
\end{IEEEeqnarray}
where the equality (a) follows from the definitions in \eqref{eq:app_check_Y}, \eqref{eq:app_def_check_X}, and \eqref{eq:app_def_check_y}. \cref{prop:layered_lb} is then proved by repeating the proof in Appendix \ref{app:proof_layered} with the caveat that the mutual information $I(\mathbf{Q};\bar{\mathbf{y}}|\hat{\mathbf{h}})$ is replaced with the lower bound in \eqref{eq:app_layer1_mi_lb}, and the mutual information $I(\check{\mathbf{X}};\check{\mathbf{y}}|\hat{\mathbf{h}},\pmb{\uptheta}_{\tau+1},\ldots,\pmb{\uptheta}_{\ell})$ is replaced with the lower bound in \eqref{eq:app_layer2_mi_lb}.

\bibliographystyle{IEEEtran}
\bibliography{IEEEabrv,myBib}

\end{document}

%% file: modelFigConf.tex
\begin{tikzpicture}[>=latex,
	ant/.style={%
		draw,
		fill,
		regular polygon,
		regular polygon sides=3,
		shape border rotate=180,
	},
]

\tikzset{ 
	wall/.pic={
		\draw[thick] (0,0) -- (-0.75cm,1cm) -- (-0.75cm,2.6cm) -- (0,1.6cm) -- (0,0);
		\draw[thick] (0,0) -- (0.15cm,0) -- (0.15cm,1.6cm) -- (-0.6cm,2.6cm) -- (-0.75cm,2.6cm);
		\draw[thick] (0.15cm,1.6cm) -- (0.0cm,1.6cm);
		\draw (0,0.2cm) -- (-0.75cm,1.2cm);
		\draw (0,0.4cm) -- (-0.75cm,1.4cm);
		\draw (0,0.6cm) -- (-0.75cm,1.6cm);
		\draw (0,0.8cm) -- (-0.75cm,1.8cm);
		\draw (0,1cm) -- (-0.75cm,2cm);
		\draw (0,1.2cm) -- (-0.75cm,2.2cm);
		\draw (0,1.4cm) -- (-0.75cm,2.4cm);
		\draw (-0.37cm,0.5cm) -- (-0.37cm,0.7cm);
		\draw (-0.37cm,0.9cm) -- (-0.37cm,1.1cm);
		\draw (-0.37cm,1.3cm) -- (-0.37cm,1.5cm);
		\draw (-0.37cm,1.7cm) -- (-0.37cm,1.9cm);
		\draw (-0.19cm,0.45cm) -- (-0.19cm,0.65cm);
		\draw (-0.19cm,0.85cm) -- (-0.19cm,1.05cm);
		\draw (-0.19cm,1.25cm) -- (-0.19cm,1.45cm);
		\draw (-0.19cm,1.65cm) -- (-0.19cm,1.85cm);
		\draw (-0.56cm,0.95cm) -- (-0.56cm,1.15cm);
		\draw (-0.56cm,1.35cm) -- (-0.56cm,1.55cm);
		\draw (-0.56cm,1.75cm) -- (-0.56cm,1.95cm);
		\draw (-0.56cm,2.15cm) -- (-0.56cm,2.35cm);
	}
}

\node (EN) [thick,draw,minimum width=1cm,minimum height=1cm, font=\small] at (0,0) {TX};
\node (Ant1) [ant, above=0.3cm of EN,scale=0.5] {};
\path[draw] (EN.north) -- (Ant1);

\node (user) [thick,draw,minimum width=1cm,minimum height=1cm,right = 5cm of EN.center, font=\small]  {RX};
\node (Ant2) [ant, above =0.3cm of user,scale=0.5, xshift=-6mm] {};
\path[draw] ($(user.north)-(3mm,0mm)$) -- (Ant2);
\node (Ant3) [ant, above =0.3cm of user,scale=0.5, xshift=6mm] {};
\path[draw] ($(user.north)+(3mm,0mm)$) -- (Ant3);

\node (IRS) [thick,draw,minimum width=1.5cm,minimum height=1.5cm, font=\small] at ($(EN)!0.5!(user)+(0,2.7cm)$) {};
\node (R1) [thick,fill=blue!50,draw,minimum width=0.2cm,minimum height=0.2cm] at ([yshift=5.5mm,xshift=-5.5mm]IRS.center){};
\node (R2) [thick,fill=blue!50,draw,minimum width=0.2cm,minimum height=0.2cm] at ([yshift=-5.5mm,xshift=-5.5mm]IRS.center){};
\node (R3) [thick,fill=blue!50,draw,minimum width=0.2cm,minimum height=0.2cm] at ([yshift=5.5mm,xshift=+5.5mm]IRS.center){};
\node (R4) [thick,fill=blue!50,draw,minimum width=0.2cm,minimum height=0.2cm] at ([yshift=-5.5mm,xshift=+5.5mm]IRS.center){};
\node (R5) [thick,fill=blue!50,draw,minimum width=0.2cm,minimum height=0.2cm] at ([yshift=+5.5mm,xshift=+1.8mm]IRS.center){};
\node (R6) [thick,fill=blue!50,draw,minimum width=0.2cm,minimum height=0.2cm] at ([yshift=+5.5mm,xshift=-1.8mm]IRS.center){};
\node (R7) [thick,fill=blue!50,draw,minimum width=0.2cm,minimum height=0.2cm] at ([yshift=-5.5mm,xshift=+1.8mm]IRS.center){};
\node (R8) [thick,fill=blue!50,draw,minimum width=0.2cm,minimum height=0.2cm] at ([yshift=-5.5mm,xshift=-1.8mm]IRS.center){};
\node (R9) [thick,fill=blue!50,draw,minimum width=0.2cm,minimum height=0.2cm] at ([yshift=2mm,xshift=-5.5mm]IRS.center){};
\node (R10) [thick,fill=blue!50,draw,minimum width=0.2cm,minimum height=0.2cm] at ([yshift=-2mm,xshift=-5.5mm]IRS.center){};
\node (R11) [thick,fill=blue!50,draw,minimum width=0.2cm,minimum height=0.2cm] at ([yshift=2mm,xshift=+5.5mm]IRS.center){};
\node (R12) [thick,fill=blue!50,draw,minimum width=0.2cm,minimum height=0.2cm] at ([yshift=-2mm,xshift=+5.5mm]IRS.center){};
\node (R13) [thick,fill=blue!50,draw,minimum width=0.2cm,minimum height=0.2cm] at ([yshift=+2mm,xshift=+1.8mm]IRS.center){};
\node (R14) [thick,fill=blue!50,draw,minimum width=0.2cm,minimum height=0.2cm] at ([yshift=+2mm,xshift=-1.8mm]IRS.center){};
\node (R15) [thick,fill=blue!50,draw,minimum width=0.2cm,minimum height=0.2cm] at ([yshift=-2mm,xshift=+1.8mm]IRS.center){};
\node (R16) [thick,fill=blue!50,draw,minimum width=0.2cm,minimum height=0.2cm] at ([yshift=-2mm,xshift=-1.8mm]IRS.center){};
\node [font=\small, above=0.0mm of IRS] {RIS};

\node (building)  at (2.1,0.35) {{\fontsize{50}{60} \faBuildingO}};
\node (car)  at (4,-0.3) {{\huge \faCar}};
\node (tree)  at (3.5,1) {{\huge \faTree}};

\coordinate (A) at ($(Ant1)+(2mm,0)$);
\path[draw,->,line width=0.5mm,red] (A) -- node[above,font=\small,pos=0.4] {$\mathbf{g}(t)$} coordinate[pos=0.2] (P1) coordinate[pos=0.25] (P2) ($(IRS)+(0,0)$);
\draw [red,line width=0.5mm] ($(P1)!0.2cm!270:(A)$) -- ($(P1)!0.2cm!90:(A)$);
\draw [red,line width=0.5mm] ($(P2)!0.2cm!270:(A)$) -- ($(P2)!0.2cm!90:(A)$);
\coordinate (B) at ($(Ant2)+(-2mm,0mm)$);
\path[draw,->,line width=0.5mm,red] ($(IRS)+(0,0)$) --  coordinate[pos=0.75] (P3) coordinate[pos=0.8] (P4) (B);
\node [font=\small,red] at ($(IRS)+(19mm,-5mm)$) {Wireless Link};
\path[draw,->,line width=0.5mm,red] ($(IRS)+(0,0)$) -- ($(tree.north)-(2mm,3mm)$) --  node[above,font=\small,pos=0.4] {$\mathbf{H}(t)$} ($(B)-(0,3mm)$);
\path[draw,->,line width=0.5mm,red] ($(IRS)+(0,0)$) -- ($(car.north)+(2mm,-1mm)$) -- ($(B)-(0,5mm)$);
\draw[draw,line width=0.5mm,red] ($(B)+(-2.5mm,4mm)$) to[out=120,in=250] ($(B)-(2.5mm,8mm)$);
\draw[draw,line width=0.5mm,red] ($(B)+(-3.5mm,4mm)$) to[out=120,in=250] ($(B)-(3.5mm,8mm)$);

\path[draw,->,line width=0.2mm] ($(EN.west)-(15mm,0)$) -- node[above,font=\small] {$w$} node[below,font=\small] {($nR$ bits)} ($(EN.west)+(0,0)$);
\path[draw,->,line width=0.2mm] ($(user.east)-(0mm,0)$) -- node[above,font=\small] {$\hat{w}$} ($(user.east)+(5mm,0)$);

\draw[line width=0.3mm,OliveGreen,->,dashed] ($(EN.north west)+(3pt,0)$) |- node[above,font=\small,pos=0.73] {Control Link} node[below,font=\small,pos=0.73] {($\text{Rate}=1/m$)} ($(IRS.west)+(0,10pt)$);

\end{tikzpicture}

%% file: timeScales.tex
\begin{tikzpicture}[>=latex,
	ant/.style={%
		draw,
		fill,
		regular polygon,
		regular polygon sides=3,
		shape border rotate=180,
	},
]

\def\minHeight{0.8cm}
\def\minWidthA{2.5cm}
\def\minWidthB{1.9cm}
\def\minWidthC{1.2cm}
\def\vertDistA{10mm}
\def\vertDistB{5mm}

\node (cb1) [thick,draw,minimum width=\minWidthA,minimum height=\minHeight, font=\small] at (0,0) {$\mathbf{g}(1)$, $\mathbf{H}(1)$};
\node (dots1) [thick,draw,minimum width=\minWidthA,minimum height=\minHeight, font=\small, right = 0 of cb1] {$\ldots$};
\node (cbt) [thick,draw,minimum width=\minWidthA,minimum height=\minHeight, font=\small, right = 0 of dots1] {$\mathbf{g}(t)$, $\mathbf{H}(t)$};
\node (dots2) [thick,draw,minimum width=\minWidthA,minimum height=\minHeight, font=\small, right = 0 of cbt] {$\ldots$};
\node (cbn) [thick,draw,minimum width=\minWidthA,minimum height=\minHeight, font=\small, right = 0 of dots2] {$\mathbf{g}(n/T)$, $\mathbf{H}(n/T)$};

\node (sbi) [thick,draw,minimum width=\minWidthB,minimum height=\minHeight, font=\small, below = \vertDistA of cbt] {$\mathbf{s}_i(t)$, $\pmb{\uptheta}_i(t)$};
\node (dots3) [thick,draw,minimum width=\minWidthB,minimum height=\minHeight, font=\small, left = 0mm of sbi] {$\ldots$};
\node (dots4) [thick,draw,minimum width=\minWidthB,minimum height=\minHeight, font=\small, right = 0mm of sbi] {$\ldots$};
\node (sb1) [thick,draw,minimum width=\minWidthB,minimum height=\minHeight, font=\small, left = 0mm of dots3] {$\mathbf{s}_1(t)$, $\pmb{\uptheta}_1(t)$};
\node (sbell) [thick,draw,minimum width=\minWidthB,minimum height=\minHeight, font=\small, right = 0mm of dots4] {$\mathbf{s}_\ell(t)$, $\pmb{\uptheta}_\ell(t)$};

\node (dots5) [thick,draw,minimum width=\minWidthC,minimum height=\minHeight, font=\small, below = \vertDistB of sbi] {$\ldots$};
\node (s1) [thick,draw,minimum width=\minWidthC,minimum height=\minHeight, font=\small, left = 0 of dots5] {$\mathrm{s}_{i,1}(t)$};
\node (sm) [thick,draw,minimum width=\minWidthC,minimum height=\minHeight, font=\small, right = 0 of dots5] {$\mathrm{s}_{i,m}(t)$};

\draw[] (cbt.south west) -- (sb1.north west);
\draw[] (cbt.south east) -- (sbell.north east);
\draw[] (sbi.south west) -- (s1.north west);
\draw[] (sbi.south east) -- (sm.north east);

\path [draw,<->] ($(cbt.north west)+(0,0.15cm)$) -- node[midway,above]{{\small coherence block ($T$ symbols)}} ($(cbt.north east)+(0,0.15cm)$);
\path [draw,<->] ($(sbi.north west)+(0,0.15cm)$) -- node[midway,above]{{\small sub-block ($m$ symbols)}} ($(sbi.north east)+(0,0.15cm)$);
\path [draw,<->] ($(s1.south west)-(0,0.15cm)$) -- node[midway,below]{{\small symbol}} ($(s1.south east)-(0,0.15cm)$);
\path [draw,<->] ($(cb1.north west)+(0,1cm)$) -- node[midway,above]{{\small coding slot ($n$ symbols)}} ($(cbn.north east)+(0,1cm)$);

\end{tikzpicture}

%% file: training.tex
\begin{tikzpicture}[>=latex,
	ant/.style={%
		draw,
		fill,
		regular polygon,
		regular polygon sides=3,
		shape border rotate=180,
	},
]

\def\minHeight{0.8cm}
\def\minWidthA{1.3cm}

\node (pilot1) [thick,draw,minimum width=\minWidthA,minimum height=\minHeight, font=\small] at (0,0) {$\bar{\bm{X}}_1$};
\node (dots1) [thick,draw,minimum width=\minWidthA,minimum height=\minHeight, font=\small, right = 0 of pilot1] {$\ldots$};
\node (pilotsTau) [thick,draw,minimum width=\minWidthA,minimum height=\minHeight, font=\small, right = 0 of dots1] {$\bar{\bm{X}}_\tau$};
\node (data1) [thick,draw,minimum width=\minWidthA,minimum height=\minHeight, font=\small, right = 0 of pilotsTau] {$\bar{\mathbf{X}}_{\tau+1}(t)$};
\node (dots2) [thick,draw,minimum width=\minWidthA,minimum height=\minHeight, font=\small, right = 0 of data1] {$\ldots$};
\node (dataEll) [thick,draw,minimum width=\minWidthA,minimum height=\minHeight, font=\small, right = 0 of dots2] {$\bar{\mathbf{X}}_{\ell}(t)$};

\path [draw,<->] ($(pilot1.south west)-(0,0.15cm)$) -- node[midway,below]{{\small pilot sub-blocks}} ($(pilotsTau.south east)-(0,0.15cm)$);
\path [draw,<->] ($(data1.south west)-(0,0.15cm)$) -- node[midway,below]{{\small data sub-blocks}} ($(dataEll.south east)-(0,0.15cm)$);
\path [draw,<->] ($(pilot1.north west)+(0,0.15cm)$) -- node[midway,above]{{\small coherence block $t$}} ($(dataEll.north east)+(0,0.15cm)$);

\end{tikzpicture}

%% file: workspace_08-09-2020-1737.tex
\definecolor{mycolor1}{rgb}{0.30100,0.74500,0.93300}%
\definecolor{mycolor2}{rgb}{0.00000,0.44700,0.74100}%
\definecolor{mycolor3}{rgb}{0.85000,0.32500,0.09800}%
\definecolor{mycolor4}{rgb}{0.92900,0.69400,0.12500}%
\begin{tikzpicture}

\begin{axis}[%
width=4.521in,
height=3.566in,
at={(0.758in,0.481in)},
scale only axis,
xmin=-20,
xmax=40,
xlabel style={font=\color{white!15!black}},
xlabel={$P$ [dB]},
ymin=0,
ymax=2.1,
ylabel style={font=\color{white!15!black}},
ylabel={Rate [bits per channel use},
axis background/.style={fill=white},
axis x line*=bottom,
axis y line*=left,
xmajorgrids,
xminorgrids,
ymajorgrids,
yminorgrids,
legend pos= north west,
legend style={legend cell align=left, align=left, draw=white!15!black}
]
\node[anchor=west,color=mycolor2] (source) at (axis cs:-20,1.7){optimal signalling w/ CSIT};
\node (destination) at (axis cs:12,1.7){};
\draw[-latex,color=mycolor2](source)--(destination);

\node[anchor=west,color=mycolor1] (source2) at (axis cs:5,0.25){uniform signalling (no CSIT)};
\node (destination2) at (axis cs:-5,0.25){};
\draw[-latex,color=mycolor1](source2)--(destination2);

\node[anchor=west,color=mycolor3] (source3) at (axis cs:17,1.3){max-SNR w/ CSIT};
\node (destination3) at (axis cs:15,1){};
\draw[-latex,color=mycolor3](source3)--(destination3);

\node[anchor=west,color=mycolor4] (source4) at (axis cs:15,0.7){max-SNR w/o CSIT};
\node (destination4) at (axis cs:15,0.95){};
\draw[-latex,color=mycolor4](source4)--(destination4);

\addplot [color=black,  line width=1.5pt, draw=none]
table[row sep=crcr]{%
	0	1\\
	15	2\\
};
\addlegendentry{exact}
\addplot [color=black, dotted, line width=1.5pt, draw=none]
table[row sep=crcr]{%
	0	1\\
	15	2\\
};
\addlegendentry{lower bound}

\addplot [color=mycolor1, line width=1.5pt]
  table[row sep=crcr]{%
40	2.03620953938625\\
35	2.03620035195809\\
30	2.03588007201018\\
25	2.03264087264202\\
20	2.00708854523959\\
15	1.88994993804358\\
10	1.58889659402432\\
5	1.14011309737426\\
0	0.644707385904214\\
-5	0.257154136807841\\
-10	0.0836041873073794\\
-15	0.0428198804523763\\
-20	0.0369561517274596\\
};

\addplot [color=mycolor1, dotted, line width=1.5pt]
  table[row sep=crcr]{%
40	2.04269611053926\\
35	2.04268298038942\\
30	2.0422514315663\\
25	2.03817400380475\\
20	2.00812610170824\\
15	1.87910083318573\\
10	1.56389501340848\\
5	1.10546697044832\\
0	0.610997171614365\\
-5	0.239597123296242\\
-10	0.0824594874983373\\
-15	0.0480335862861666\\
-20	0.0432907593303453\\
};

\addplot [color=mycolor2, line width=1.5pt]
  table[row sep=crcr]{%
40	2.03620953938625\\
35	2.03620035199783\\
30	2.03588039439183\\
25	2.03269793533148\\
20	2.00775214302204\\
15	1.89373167628906\\
10	1.59458420618437\\
5	1.15016773132545\\
0	0.677541215470057\\
-5	0.304297044150652\\
-10	0.102129504083218\\
-15	0.0459354812027234\\
-20	0.0373253759926774\\
};

\addplot [color=mycolor2, dotted, line width=1.5pt]
  table[row sep=crcr]{%
40	2.04269613950711\\
35	2.04268300942165\\
30	2.04226068171483\\
25	2.03833579406959\\
20	2.00952732880336\\
15	1.88442119773489\\
10	1.57016525887352\\
5	1.11686305669409\\
0	0.644108909605266\\
-5	0.278173126275213\\
-10	0.0957881416251044\\
-15	0.0500853135602827\\
-20	0.0435247546430397\\
};

\addplot [color=mycolor3, line width=1.5pt]
  table[row sep=crcr]{%
40	1.03620958985152\\
35	1.03620958985152\\
30	1.03620937845372\\
25	1.0361095090912\\
20	1.03359131718141\\
15	1.00071849151117\\
10	0.852779557892304\\
5	0.596920635887605\\
0	0.339924339886131\\
-5	0.150187587124727\\
-10	0.061536409458639\\
-15	0.0398398750748453\\
-20	0.0366273903723774\\
};

\addplot [color=mycolor3, dotted, line width=1.5pt]
  table[row sep=crcr]{%
40	1.04269610755903\\
35	1.04269610755903\\
30	1.04269517440063\\
25	1.04241668110353\\
20	1.03688842744392\\
15	0.986326679066242\\
10	0.816864560794653\\
5	0.555774408776718\\
0	0.304806281188536\\
-5	0.134079347270346\\
-10	0.0619970506074559\\
-15	0.0454111230358574\\
-20	0.043009489560062\\
};

\addplot [color=mycolor4, line width=1.5pt]
  table[row sep=crcr]{%
40	1.0362095211075\\
35	1.0362095211075\\
30	1.03579290367029\\
25	1.03250238099557\\
20	1.02360189427994\\
15	0.937364240471146\\
10	0.760221033962192\\
5	0.519265195264788\\
0	0.281686246895663\\
-5	0.12794230255\\
-10	0.0560323384579239\\
-15	0.0392737720972516\\
-20	0.0365674286396521\\
};

\addplot [color=mycolor4, dotted, line width=1.5pt]
  table[row sep=crcr]{%
40	1.04269610755903\\
35	1.04266808797163\\
30	1.04265118739051\\
25	1.03966386749607\\
20	1.02258029003066\\
15	0.922834773758324\\
10	0.729002024769265\\
5	0.465240043865919\\
0	0.261081165525107\\
-5	0.116588317093883\\
-10	0.0584606763684004\\
-15	0.0449669436467904\\
-20	0.0429506869104597\\
};

\end{axis}

\begin{axis}[%
width=5.833in,
height=4.375in,
at={(0in,0in)},
scale only axis,
xmin=0,
xmax=1,
ymin=0,
ymax=1,
axis line style={draw=none},
ticks=none,
axis x line*=bottom,
axis y line*=left,
legend style={legend cell align=left, align=left, draw=white!15!black}
]
\end{axis}
\end{tikzpicture}%

%% file: workspace_08-11-2020-1011.tex
\definecolor{mycolor1}{rgb}{0.49400,0.18400,0.55600}%
\definecolor{mycolor2}{rgb}{0.00000,0.44700,0.74100}%
\definecolor{mycolor3}{rgb}{0.30100,0.74500,0.93300}%
\definecolor{mycolor4}{rgb}{0.46600,0.67400,0.18800}%
\definecolor{mycolor5}{rgb}{0.85000,0.32500,0.09800}%
\definecolor{mycolor6}{rgb}{0.92900,0.69400,0.12500}%
\begin{tikzpicture}

\begin{axis}[%
width=4.521in,
height=3.566in,
at={(0.758in,0.481in)},
scale only axis,
xmin=0,
xmax=20,
xlabel style={font=\color{white!15!black}},
xlabel={$\tau$},
ymin=0,
ymax=6,
ylabel style={font=\color{white!15!black}},
ylabel={Rate [bit per channel use]},
axis background/.style={fill=white},
axis x line*=bottom,
axis y line*=left,
xmajorgrids,
xminorgrids,
ymajorgrids,
yminorgrids,
legend style={legend cell align=left, align=left, draw=white!15!black}
]
\addplot [color=mycolor1, dashed, line width=2pt]
  table[row sep=crcr]{%
0	5.9751355787369\\
19	5.9751355787369\\
};
\addlegendentry{optimal signalling w/ perfect CSI}

\addplot [color=mycolor2, line width=1.5pt, mark size=4.0pt, mark=o, mark options={solid, mycolor2}]
  table[row sep=crcr]{%
0	1.06761699217582\\
1	3.24373623926427\\
2	3.90446656210081\\
3	4.33900228643196\\
4	4.75943909329196\\
5	4.49556249411795\\
6	4.15631690414117\\
7	3.84660584368349\\
8	3.56607503239038\\
9	3.2757649855092\\
10	2.96497945740444\\
11	2.66583664328798\\
12	2.37108932020078\\
13	2.08918073676311\\
14	1.78291411039986\\
15	1.47881871881715\\
16	1.20219748482411\\
17	0.928713300595468\\
18	0.618363689385379\\
19	0.314576470416051\\
};
\addlegendentry{optimal signalling w/ imperfect CSI}

\addplot [color=mycolor3, line width=1.5pt, mark size=4.0pt, mark=x, mark options={solid, mycolor3}]
  table[row sep=crcr]{%
0	0.942211244806888\\
1	1.9865261134723\\
2	2.96083592813296\\
3	3.93971869796031\\
4	4.75943909329196\\
5	4.49556249411795\\
6	4.15631690414117\\
7	3.84660584368349\\
8	3.56607503239038\\
9	3.2757649855092\\
10	2.96497945740444\\
11	2.66583664328798\\
12	2.37108932020078\\
13	2.08918073676311\\
14	1.78291411039986\\
15	1.47881871881715\\
16	1.20219748482411\\
17	0.928713300595468\\
18	0.618363689385379\\
19	0.314576470416051\\
};
\addlegendentry{uniform signalling (no CSIT)}

\addplot [color=mycolor4, dotted, line width=2pt]
  table[row sep=crcr]{%
0	1.97513558493579\\
19	1.97513558493579\\
};
\addlegendentry{max-SNR w/ perfect CSI}

\addplot [color=mycolor5, line width=1.5pt, mark size=2.8pt, mark=square, mark options={solid, mycolor5}]
  table[row sep=crcr]{%
0	0.94221123455489\\
1	1.86898081666007\\
2	1.75190316094033\\
3	1.54570208780169\\
4	1.55943909662981\\
5	1.4955624870548\\
6	1.35631691507267\\
7	1.24660584926249\\
8	1.16607505042269\\
9	1.07576498307674\\
10	0.964979461424174\\
11	0.865836645674068\\
12	0.771089320260382\\
13	0.689180736825699\\
14	0.582914111249222\\
15	0.478818717505844\\
16	0.402197485086371\\
17	0.328713300711698\\
18	0.218363691303158\\
19	0.114576471285286\\
};
\addlegendentry{max-SNR w/ imperfect CSI}

\addplot [color=mycolor6, line width=1.5pt, mark size=4.0pt, mark=asterisk, mark options={solid, mycolor6}]
  table[row sep=crcr]{%
0	0.94221123455489\\
1	1.05581497638561\\
2	1.14785600825266\\
3	1.18690911780941\\
4	1.55943909662981\\
5	1.4955624870548\\
6	1.35631691507267\\
7	1.24660584926249\\
8	1.16607505042269\\
9	1.07576498307674\\
10	0.964979461424174\\
11	0.865836645674068\\
12	0.771089320260382\\
13	0.689180736825699\\
14	0.582914111249222\\
15	0.478818717505844\\
16	0.402197485086371\\
17	0.328713300711698\\
18	0.218363691303158\\
19	0.114576471285286\\
};
\addlegendentry{max-SNR w/o CSIT}

\end{axis}

\begin{axis}[%
width=5.833in,
height=4.375in,
at={(0in,0in)},
scale only axis,
xmin=0,
xmax=1,
ymin=0,
ymax=1,
axis line style={draw=none},
ticks=none,
axis x line*=bottom,
axis y line*=left,
legend style={legend cell align=left, align=left, draw=white!15!black}
]
\end{axis}
\end{tikzpicture}%

%% file: workspace_08-12-2020-1035.tex
\definecolor{mycolor1}{rgb}{0.00000,0.44700,0.74100}%
\definecolor{mycolor2}{rgb}{0.30100,0.74500,0.93300}%
\definecolor{mycolor3}{rgb}{0.85000,0.32500,0.09800}%
\definecolor{mycolor4}{rgb}{0.92900,0.69400,0.12500}%
\begin{tikzpicture}

\begin{axis}[%
width=4.521in,
height=3.566in,
at={(0.758in,0.481in)},
scale only axis,
xmin=1,
xmax=10,
xlabel style={font=\color{white!15!black}},
xlabel={$\ell$},
ymin=0.4,
ymax=2.2,
ylabel style={font=\color{white!15!black}},
ylabel={Rate [bit per channel use]},
axis background/.style={fill=white},
axis x line*=bottom,
axis y line*=left,
xmajorgrids,
xminorgrids,
ymajorgrids,
yminorgrids,
legend style={at={(0.01,0.99)}, anchor=north west, legend cell align=left, align=left, draw=white!15!black}
]
\addplot [color=mycolor1, line width=1.5pt, mark size=4.0pt, mark=o, mark options={solid, mycolor1}]
  table[row sep=crcr]{%
1	0.653049713485062\\
2	1.02429415884881\\
3	1.36572554513174\\
4	1.53644123827321\\
5	1.63887065415809\\
6	1.76799787216413\\
7	1.89428343446157\\
8	1.98899760618465\\
9	2.06266418419149\\
10	2.14813614825286\\
};
\addlegendentry{optimal signalling w/ CSIT}

\addplot [color=mycolor2, line width=1.5pt, mark size=4.0pt, mark=x, mark options={solid, mycolor2}]
  table[row sep=crcr]{%
1	0.520294593058649\\
2	0.520294593058649\\
3	0.676814636361034\\
4	0.821880037226975\\
5	0.98625604467237\\
6	1.16879136933837\\
7	1.50273176057791\\
8	1.75318705400756\\
9	1.94798561556396\\
10	2.10382446480907\\
};
\addlegendentry{uniform signalling (no CSIT)}

\addplot [color=mycolor3, line width=1.5pt, mark size=2.8pt, mark=square, mark options={solid, mycolor3}]
  table[row sep=crcr]{%
1	0.520294559501231\\
2	0.564011330055804\\
3	0.752015106741072\\
4	0.846016995083706\\
5	0.913230319964362\\
6	1.01470035551596\\
7	1.08717895233853\\
8	1.14153789995545\\
9	1.18381708143528\\
10	1.21764042661915\\
};
\addlegendentry{max-SNR w/ CSIT}

\addplot [color=mycolor4, line width=1.5pt, mark size=4.0pt, mark=asterisk, mark options={solid, mycolor4}]
  table[row sep=crcr]{%
1	0.520294559501231\\
2	0.520294559501231\\
3	0.520294559501231\\
4	0.520294559501231\\
5	0.520294559501231\\
6	0.520294559501231\\
7	0.531063915333673\\
8	0.600417704580624\\
9	0.667130782867361\\
10	0.720501245496749\\
};
\addlegendentry{max-SNR w/o CSIT }

\end{axis}
\end{tikzpicture}%

%% file: workspace_08-12-2020-1417.tex
\definecolor{mycolor1}{rgb}{0.00000,0.44700,0.74100}%
\definecolor{mycolor2}{rgb}{0.30100,0.74500,0.93300}%
\definecolor{mycolor3}{rgb}{0.85000,0.32500,0.09800}%
\definecolor{mycolor4}{rgb}{0.92900,0.69400,0.12500}%
\begin{tikzpicture}

\begin{axis}[%
width=4.521in,
height=3.566in,
at={(0.758in,0.481in)},
scale only axis,
xmin=1,
xmax=10,
xlabel style={font=\color{white!15!black}},
xlabel={$N$},
ymin=0,
ymax=6,
ylabel style={font=\color{white!15!black}},
ylabel={Rate [bit per channel use]},
axis background/.style={fill=white},
axis x line*=bottom,
axis y line*=left,
xmajorgrids,
xminorgrids,
ymajorgrids,
yminorgrids,
legend style={at={(0.97,0.5)}, anchor=east, legend cell align=left, align=left, draw=white!15!black}
]
\addplot [color=mycolor1, line width=1.5pt, mark size=4.0pt, mark=o, mark options={solid, mycolor1}]
  table[row sep=crcr]{%
1	2.98416524956927\\
2	4.64535504502078\\
3	5.27594706162315\\
4	5.52939045914021\\
5	5.61409788218236\\
6	5.61409788218236\\
7	5.61409788218236\\
8	5.61409788218236\\
9	5.61409788218236\\
10	5.61409788218236\\
};
\addlegendentry{optimal signalling w/ CSIT}

\addplot [color=mycolor2, line width=1.5pt, mark size=4.0pt, mark=x, mark options={solid, mycolor2}]
  table[row sep=crcr]{%
1	2.92311408063541\\
2	4.6439685548894\\
3	5.27555385867839\\
4	5.52924928394642\\
5	5.61406125235868\\
6	5.61406125235868\\
7	5.61406125235868\\
8	5.61406125235868\\
9	5.61406125235868\\
10	5.61406125235868\\
};
\addlegendentry{uniform signalling (no CSIT)}

\addplot [color=mycolor3, line width=1.5pt, mark size=2.8pt, mark=square, mark options={solid, mycolor3}]
  table[row sep=crcr]{%
1	0.764477579644539\\
2	0.781511565634207\\
3	0.790010139038615\\
4	0.857538671679885\\
5	0.862130520409829\\
6	0.862130520409829\\
7	0.862130520409829\\
8	0.862130520409829\\
9	0.862130520409829\\
10	0.862130520409829\\
};
\addlegendentry{max-SNR w/ CSIT}

\addplot [color=mycolor4, line width=1.5pt, mark size=4.0pt, mark=asterisk, mark options={solid, mycolor4}]
  table[row sep=crcr]{%
1	0.631272855871908\\
2	0.749365323759331\\
3	0.786115871740393\\
4	0.856112276359469\\
5	0.861940883466012\\
6	0.861940883466012\\
7	0.861940883466012\\
8	0.861940883466012\\
9	0.861940883466012\\
10	0.861940883466012\\
};
\addlegendentry{max-SNR w/o CSIT }

\end{axis}
\end{tikzpicture}%

%% file: workspace_09-11-2020-1314.tex
\definecolor{mycolor1}{rgb}{0.00000,0.44700,0.74100}%
\definecolor{mycolor2}{rgb}{0.46600,0.67400,0.18800}%
\definecolor{mycolor3}{rgb}{0.85000,0.32500,0.09800}%
\begin{tikzpicture}

\begin{axis}[%
width=4.521in,
height=3.566in,
at={(0.758in,0.481in)},
scale only axis,
xmin=-20,
xmax=40,
xlabel style={font=\color{white!15!black}},
xlabel={$P$ [dB]},
ymin=0,
ymax=3.5,
ylabel style={font=\color{white!15!black}},
ylabel={Rate [bits per channel use]},
axis background/.style={fill=white},
axis x line*=bottom,
axis y line*=left,
xmajorgrids,
ymajorgrids,
legend pos= north west,
legend style={legend cell align=left, align=left, draw=white!15!black}
]

\node (A) at (axis cs:32,3.05){};
\draw[thick,draw=black] (A) ellipse (7pt and 30pt);
\node[color=mycolor1] at (axis cs:22,3.05){optimal signalling };
\node (B) at (axis cs:13.5,2.3){};
\draw[thick,draw=black] (B) ellipse (3pt and 10pt);
\node[color=mycolor2] at (axis cs:23,2.43){layered encoding};
\node (C) at (axis cs:10,1.83){};
\draw[thick,draw=black] (C) ellipse (5pt and 15pt);
\node[color=mycolor3] at (axis cs:17,1.7){max-SNR};

\addplot [color=black,  line width=1.5pt, draw=none]
table[row sep=crcr]{%
	0	1\\
	15	2\\
};
\addlegendentry{4-ASK}
\addplot [color=black, dashed, line width=1.5pt, draw=none]
table[row sep=crcr]{%
	0	1\\
	15	2\\
};
\addlegendentry{QPSK}

\addplot [color=mycolor1, dashed, line width=1.5pt]
  table[row sep=crcr]{%
40	2.81712923705391\\
35	2.81712923705391\\
30	2.81712923705391\\
25	2.81712587900649\\
20	2.8170018501704\\
15	2.81469765880713\\
10	2.78946211306049\\
5	2.6212955836507\\
0	2.05128933331701\\
-5	1.14860911835498\\
-10	0.375690803917051\\
-15	0.0653345256623635\\
-20	0.00566021572404791\\
};

\addplot [color=mycolor2, dashed, line width=1.5pt]
  table[row sep=crcr]{%
40	2.33461197101248\\
35	2.33461197101248\\
30	2.33461197101248\\
25	2.33461197101248\\
20	2.3346037678994\\
15	2.33223519981874\\
10	2.29961460030052\\
5	2.11071791635327\\
0	1.54182765784237\\
-5	0.737151342190781\\
-10	0.197950658971237\\
-15	0.0205391364838098\\
-20	-0.0109945728577262\\
};

\addplot [color=mycolor3, dashed, line width=1.5pt]
  table[row sep=crcr]{%
40	1.87712936390213\\
35	1.87712936390213\\
30	1.87712936390213\\
25	1.87712936390213\\
20	1.87712936390213\\
15	1.87712936390213\\
10	1.87709411586835\\
5	1.87071197672483\\
0	1.73879392638502\\
-5	1.13264007741748\\
-10	0.376025533855502\\
-15	0.0653653497492084\\
-20	0.00566052897789567\\
};

\addplot [color=mycolor1, line width=1.5pt]
  table[row sep=crcr]{%
40	3.28712926753337\\
35	3.28712914793895\\
30	3.28710899742673\\
25	3.28650477145789\\
20	3.27754761122632\\
15	3.21675156872805\\
10	2.9434826552958\\
5	2.32610116411061\\
0	1.51143473329274\\
-5	0.786835644944597\\
-10	0.301255118479412\\
-15	0.0624529111004639\\
-20	0.00613123592346201\\
};

\addplot [color=mycolor2, line width=1.5pt]
  table[row sep=crcr]{%
40	2.33461197101248\\
35	2.33461197101248\\
30	2.33461179905209\\
25	2.33456577511105\\
20	2.33059102705889\\
15	2.29693655192924\\
10	2.13561069632444\\
5	1.76260184040924\\
0	1.14593973843627\\
-5	0.519859765235438\\
-10	0.131636272729416\\
-15	0.00992841683630288\\
-20	-0.0121680807727289\\
};

\addplot [color=mycolor3, line width=1.5pt]
  table[row sep=crcr]{%
40	1.87712936390213\\
35	1.87712936390213\\
30	1.87712936390213\\
25	1.87712932559516\\
20	1.87706037931256\\
15	1.87208259064959\\
10	1.78906255639151\\
5	1.43861667335351\\
0	0.932842635262156\\
-5	0.47265569024716\\
-10	0.132718571182971\\
-15	0.0196203477883904\\
-20	0.000113530752059715\\
};

\end{axis}

\begin{axis}[%
width=5.833in,
height=4.375in,
at={(0in,0in)},
scale only axis,
xmin=0,
xmax=1,
ymin=0,
ymax=1,
axis line style={draw=none},
ticks=none,
axis x line*=bottom,
axis y line*=left,
legend style={legend cell align=left, align=left, draw=white!15!black}
]
\end{axis}
\end{tikzpicture}%

%% file: N=2_K=2_A=2_P=40_numInputs=2_vs_DT.tex
\definecolor{mycolor1}{rgb}{0.4940,0.1840,0.5560}%
\definecolor{mycolor2}{rgb}{0.4660,0.6740,0.1880}%
\definecolor{mycolor3}{rgb}{0.3010, 0.7450, 0.9330}%
\begin{tikzpicture}

\begin{axis}[%
width=4.521in,
height=3.0in,
at={(0.758in,0.481in)},
scale only axis,
xmin=1,
xmax=7,
xlabel style={font=\color{white!15!black}},
xlabel={$m$},
ymin=0.8,
ymax=3,
ylabel style={font=\color{white!15!black}},
ylabel={Rate [bits per channel use]},
axis background/.style={fill=white},
xmajorgrids,
ymajorgrids,
xticklabels={,1,,2,,3,,4,,5,,6,,7}, 
ytick={0.8,1,1.2,1.4,1.6,1.8,2,2.2,2.4,2.6,2.8,3},
legend style={legend cell align=left, align=left, draw=white!15!black}
]
\addplot [color=mycolor1, line width=2.0pt, mark size=4.0pt, mark=o, mark options={solid, mycolor1}]
  table[row sep=crcr]{%
1	3\\
2	2\\
3	1.66666666666667\\
4	1.5\\
5	1.4\\
6	1.33333333333333\\
7	1.28571428571429\\
};
\addlegendentry{optimal signalling}

\addplot [color=mycolor2, line width=2.0pt, mark size=2.8pt, mark=square, mark options={solid, mycolor2}]
  table[row sep=crcr]{%
1	1\\
2	1\\
3	1\\
4	1\\
5	1\\
6	1\\
7	1\\
};
\addlegendentry{max-SNR} 

\addplot [color=mycolor3, line width=2.0pt, mark size=4.0pt, mark=asterisk, mark options={solid, mycolor3}]
  table[row sep=crcr]{%
1	1\\
2	1.5\\
3	1.33333333333333\\
4	1.25\\
5	1.2\\
6	1.16666666666667\\
7	1.14285714285714\\
};
\addlegendentry{layered encoding}

\end{axis}

\end{tikzpicture}%